\documentstyle[11pt,aaspp4]{article}

\lefthead{}
\righthead{}

\begin{document}
\title{Starcounts Redivivus. IV.  Density Laws Through Photometric Parallaxes}

\author{M. H. Siegel\altaffilmark{1,4}, S. R. Majewski\altaffilmark{1,2,3}, 
\\I.~N. Reid\altaffilmark{2,4},
\\ and I.~B. Thompson\altaffilmark{5}
}

\altaffiltext{1}{Department of Astronomy, University of Virginia,
Charlottesville, VA, 22903 (mhs4p@virginia.edu,srm4n@didjeridu.astro.virginia.edu)}

\altaffiltext{2}{Visiting Research Associate, The Observatories of the Carnegie
Institution of Washington, 813 Santa Barbara Street, Pasadena, CA 91101}

\altaffiltext{3}{David and Lucile Packard Foundation Fellow}

\altaffiltext{4}{Space Telescope Science Institute, 3700 San Martin Drive, Baltimore, MD  
21218 (msiegel@stsci.edu,inr@stsci.edu)}

\altaffiltext{5}{The Observatories of the Carnegie Institution of Washington,
813 Santa Barbara Street, Pasadena, CA 91101 (ian@ociw.edu)}

\begin{abstract}
In an effort to more precisely define the spatial distribution of Galactic field stars, we 
present an analysis of the photometric parallaxes of stars in seven Kapteyn Selected Areas.
Our photometry database covers $\sim14.9$ square degrees and contain over 130,000 stars, of which 
approximately 70,000
are in a color range ($0.4 \leq R-I \leq 1.5$) for which relatively unambiguous photometric 
parallaxes can be derived.  We discuss our
photometry pipeline, our method of determining photometric parallaxes and our analysis
efforts.  We also address the affects of Malmquist Bias, subgiant/giant contamination, metallicity
and binary stars upon the derived density laws.  The affect of binary stars is the most significant
of these biases -- a binary star fraction of 50\% could result in derived scale heights that are 80\%
of the actual values.

We find that while the disk-like populations of the Milky Way are easily constrained in
a simultaneous analysis of all seven fields, no good simultaneous solution for the halo is found.  
We have applied halo density laws taken from other studies and find that the Besan\c{c}on 
flattened power law halo model ($\frac{c}{a}=0.6, \rho \propto r^{-2.75}$) produces the best fit
to our data.  
With this halo, the thick disk has a scale height of 750 pc with an 8.5\% normalization to the old 
disk.  The old disk scale height is $\sim$ 280-300 pc for our early type ($5.8 \leq M_R < 6.8$) dwarfs 
and rises to $\sim$ 350 pc for our late type ($8.8 \leq M_R \leq 10.2$) dwarf stars.  Corrected for
a binary fraction of 50\%, these scale heights are 940 pc and 350-375 pc, respectively.

Even with this model, there are systematic discrepancies between the observed and 
predicted density distributions -- discrepancies only apparent at the faint magnitudes 
reached by our survey.  
Specifically, our model produces density overpredictions
in the inner Galaxy and density underpredictions in the outer Galaxy.  A possible escape from this dilemma 
is offered by modeling the stellar halo as a two-component system, as favored by studies
of BHB/RR Lyrae stars and recent analyses of the kinematics of metal-poor stars.  In this paradigm, the halo 
has a flattened inner distribution and a roughly spherical, but substructured outer distribution.  Further 
reconciliation could be provided by a flared thick disk, a structure consistent with a merger
origin for that population.
\end{abstract}


\section{Introduction}

The present stellar content of the Milky Way is a fossil record of its formation and evolution.
When a star is formed, its kinematical and chemical properties reflect the state of the 
Galaxy at that location at the time of the star's formation.  In high density chaotic regions of the Galaxy
 -- such as the Galactic midplane -- that information can be quickly scrambled.  In the more remote 
regions of the Galaxy, the story told by ancient stars is still legible.  
Since lower-mass stars have lifetimes of order a Hubble time or greater, they remain in the Galaxy as 
echoes of the its distant past.  The most sensible and efficient strategy for reading the messages contained
in old stars is to ascertain the bulk properties of stars grouped together by similar characteristics, i.e. 
stellar populations.

The division of Galactic field stars into distinct populations was greatly clarified by Baade's (1944) division of stars
into Population I and Population II.  This system was expanded into five populations
in the seminal 1957 Vatican conference (O'Connell 1958).  Even after nearly six decades of work, however, 
there is still some uncertainty on the exact characteristics of each population and, more importantly, what those 
characteristics tell us about the evolution of the Galaxy.  Indeed, there is not even a consensus on {\it how many} 
populations are required to fully describe the Milky Way and whether distinct populations remain
a sensible paradigm (see reviews in Majewski 1993, 1999a). 

This series' first contribution to the debate (Reid \& Majewski 1993, hereafter Paper I) used 
photographic starcounts data to derive an interim model for the spatial distribution of field stars toward
the Galactic poles.  
Our second (Reid et al. 1996, hereafter Paper II) investigated small but extremely deep data sets to apply 
constraints to the halo luminosity function.  A later contribution (Majewski et al. 1997, hereafter 
Paper III) investigated an anomalous starcounts signature discovered in starcount data not presented here, 
and explored the possibility of contamination of our starcount data by streams of stars tidally stripped from the 
Sagittarius dSph galaxy.

The present discussion is a more complete and sophisticated investigation of a number 
of Kapteyn Selected Areas using large area, CCD-based datasets.
\S2 details the general characteristics of this dataset, while \S3 and \S4
detail the photometry pipeline and object classification methods, respectively.  We then use
a subset of this data and the method of photometric parallax to apply stronger global constraints to 
the spatial distribution of Galactic field
stars than any previous starcounts survey has been capable of.  \S5 details our method of photometric 
parallax, \S6-8 cover the analysis of the starcounts and our attempt to fit a self-consistent model to the data in all
seven fields and \S9 discusses the implication of those results.  \S10 summarizes the primary results of this
endeavor.

\section{Observational Program}

\subsection{Overview}

The statistical foundation of starcounts rests in the Von Seeliger's (1898) Fundamental Equation of Stellar 
Statistics.  Expanded to include stellar colors, this equation may be written as:

\begin{equation}
A(m_V,S_{B-V}) = \sum A_i (m_v, S_{B-V})=\Omega \sum \int_0^{R_{max}} \Phi_i(M,S) D_i(r)r^2dr
\end{equation}

\noindent in which $A$ is the differential number of counts at any particular magnitude and color, $A_i$ is the contribution
to those counts from population $i$, $\Omega$ is the solid angle observed, $\Phi_i$ is the luminosity function
of population $i$ and $D_i$ is the density distribution of population $i$.  In this formulation, the number of 
counts at a given color and magnitude is the sum over the stellar populations of the convolution
of the luminosity and density distribution functions.  Deriving scientific results from 
starcounts is difficult because the starcounts equation is not invertable.  Because of the
non-invertability and the vagaries of solving the non-unique convolution by trial an error, 
starcounts {\it by themselves} are in general a weak tool for exploring the Galaxy.  In 
combination with a small amount of external information, however, they can become a potent implement.

The non-uniqueness problem is perhaps demonstrated by the breadth of Galactic structure models that
have been derived from starcount studies, as listed in Table I.
While the derived structural parameters for
the thin disk occupy a fairly narrow range of values, the density law of the thick disk
\footnote{We succumb to the pressure of popular use of the term ``thick disk" in this paper, despite
potential implications for the structure and origin of this population that may inadvertently be implied
by this expression.} 
(Intermediate Population II or IPII in the Vatican Conference nomenclature) is less 
certain, with a large range of values as illustrated in Figure 1.  Although there appears
to be a cluster of results in Figure 1, this  may be a selection effect.  The 
tabulated studies have often explored similar data sets with similar limitations.  In fact, several
of them probe the same direction in the sky (the Galactic poles, especially the North Galactic Pole).  It
is perhaps encouraging that they produce similar results for one direction of sky.  It should be noted 
that there is still debate 
over whether the thick disk is, in fact, a distinct population or is just an extended tail of the old 
disk.  However, as far as the halo is concerned, the results from starcount
surveys span almost the entire range of parameter space from flattened de Vaucouleurs spheroids (Wyse 
\& Gilmore, 1989; Larsen
1996, hereafter L96) to perfectly spherical power law distributions (Ng et al. 1997, hereafter N97).

As part of a campaign to obtain photometry, spectroscopy and proper motions for stars in a large number 
of the Kapteyn Selected Areas, we have collected a set of photometric imaging data suitable for starcounts
analyses.  These data are notably superior to previous 
efforts and are of high enough quality to provide a critical check on the early results from SDSS as 
presented, for example, in Chen et al. 2001 (hereafter C01).  The specific advantages of our starcounts program are:

1.  {\it Multiple lines of sight.}  An over-reliance upon data taken in a few directions in the 
sky is characteristic of much of the prior starcounts literature.  Most studies have focused upon one or a few 
directions in the sky generally in either tiny areas to great depth (e.g, Paper I, Paper II) or 
over a large area to shallower depth (e.g. Gilmore \& Reid 1983, hereafter GR83).  Only a few programs 
have attempted to survey the Galaxy in multiple directions.  Chief among those are the 
Basel Halo Program (most recent contribution by Buser et al. 1999, hereafter B99), which has been examining program
fields with photographic plates for some four decades; the Besan\c{c}on program (e.g. Robin et al. 1996, hereafter
R96), which attempts to use population synthesis to analyze data from their own program fields
as well as previous starcount studies; the APS-POSS program (L96), the most extensive
survey to date; and the Sloan Digital Sky Survey (SDSS), which has recently published its 
first analysis (C01).

As noted in Paper I and in R96, evaluation of starcounts
in a single direction can lead to degenerate density law solutions.  In addition, surveys that 
only probe at the 
Galactic poles (in Figure 1 this includes GR83, Gilmore 1984, Robin \& Creze 1986, Yoshii et al. 1987, 
Kuijken \& Gilmore 1989, Paper I, N97)
are completely insensitive to radial terms in the population distributions.  The photometry program 
presented here consists 
of eleven Kapteyn Selected Areas plus one additional area.  
Table II lists the observed fields and Figure 2 shows an Aitoff 
projection of their distribution across the Galaxy.  These fields cover the South Galactic Pole, the 
Galactic anti-center, the $90^{\circ}-270^{\circ}$ meridional slice and locations $40^{\circ}$ above and 
below the Galactic Center.  

2.  {\it Completeness to faint magnitudes.} The average photometric depth of our data is $V\sim21.4$ and 
a substantial portion image to $V \sim 22$, a depth 
comparable to the deepest moderate area surveys conducted thus far (Paper I; Robin et al. 2000, 
hereafter R00; C01) and 
significantly deeper than the large surveys of L96, R96 or B99.  While we can not match the deepest
starcount surveys, such as those from the Hubble Deep Field and Flanking Fields (M{\' e}ndez \& Guzman 1998; 
Paper II) or SA 57 
(Paper I), those surveys have covered only a single direction of sky and/or a very small solid angle 
with a commensurately low number of stars.  Even the multi-directional HST-based survey of 
Zheng et al. (2001) has an average of only 10 stars per pointing.
Deep wide field starcounts can be extracted from galaxy count surveys (see, e.g., Phleps et al. 2000).  
Deeper surveys, however, run into two limitations.  
Star-galaxy separation grows increasingly important, but increasingly difficult, at fainter 
magnitudes.  In fact, it may be impossible to distinguish accurately between stars and galaxies 
in a ground-based survey much fainter than $V \sim 22.5$ except with data from large-aperture telescopes with fine 
plate scales observed under exceptional seeing conditions.  This problem is more significant for studies of stars
than for galaxies because the latter far outnumber the former at faint ($V>22$) magnitudes.  A second complication is the 
low sensitivity of starcounts to the parameters of the halo.  Because of the large volume over which halo stars are 
spread, stingent constraints are difficult to apply.  Thus deeper surveys will not necessarily produce superior 
results (see Paper II and Reid et al. 1998 for a discussion of both of those 
problems).

3.  {\it Linear photometry over a large dynamic range.}  Most of the previous 
surveys (all of those listed in Table I with the exception of N97 and C01) have derived their 
results from studies of photographic plates.  While photography has the distinct advantage of 
imaging a large section of sky in a single observation, the non-linearity of the medium requires 
careful calibration by photometric standards.  The fraction of counts contributed by 
each population is a function of magnitude and thus any 
error in calibration will manifest itself as a bias in the derived density laws.  Additionally, the large dynamic
range of our data  ($12 \leq V \leq 22$) provides a significant advantage over purely deep surveys, 
such as those using HST, that can only probe the most
distant stars in the Galaxy.  Gizis \& Reid (1999), for example, have shown that HST-based halo 
luminosity functions are discrepant from ground-based studies possibly because of the inability of deep 
surveys to probe the nearby Galaxy.

4.  {\it Sky coverage.}  One of the most important 
aspects of a starcounts survey is area.  The greater the 
area surveyed, the less the data are sensitive to small density fluctuations in the Galaxy.  
We observed the Selected Areas in our study over 1.5-2.5 square 
degrees each.  The only CCD-based study that exceeds this coverage is C01, which covers 279 deg$^2$.
Our survey is much more extensive than previous CCD surveys, such as 
R00 (1.4 deg$^2$), Paper II (0.0124 deg$^2$) and Zheng et al. 
($\sim 0.2$ deg $^2$).  Several photographic studies (e.g., L96, GR83) cover larger areas.

5.  {\it Homogeneity.}  Many of the difficulties listed above can be overcome by combining multiple 
data sets.  R96 and R00 approach starcounts with this method.  
Those studies, however, specifically note the
lack of homogeneity as a potential problem, with possible discrepancies in the apparent
starcounts from field to field.  Indeed, they note that there are discrepancies among different studies
of the {\it same} fields.  Our data are taken
through the same telescope with the same combination of filters, calibrated to the same standards and reduced through the same 
pipeline.

While our survey may be inferior to others in any particular aspect, 
the {\it combination} of the above advantages gives our study a view of the Galaxy that is only
matched by C01.  Even the latter surveys an area (an SDSS strip) not necessarily optimized for constraining density
laws for all stellar populations, although this sensitivity will grow as SDSS covers more of the sky.
Another comparable survey is the APS-POSS survey (L96), which covers 1440 square degrees in ninety 
directions over the magnitude interval $12 < O < 20$ and with
careful photoelectric calibration to remove the non-linearity inherent to the photographic plates.
Although APS-POSS lacks the depth of our study or C01, these shortcomings are compensated for to some extent 
by the sheer volume of stars surveyed (a staggering 2.6 million stars of the nearly $10^9$ available from the POSS).

So far, this discussion has focused on optical starcounts.  Much progress may be made in the analysis
of infrared starcounts, now available from the 2MASS program, which covers the entire sky with a 
homogenous data sample.  Because 2MASS can dramatically reduce
the effect of interstellar reddening, NIR starcounts will be valuable at low Galactic 
latitudes, where they can reach all the way to the opposite side of the Galaxy.  2MASS's primary limitation is a 
rather shallow magnitude limit (see, e.g. Beichman et al. 1998) and the extremely narrow NIR 
color range of K- and M-type stars, which limits sensitivity to spectral type.

In summary, our study has sufficient depth and number of stars to probe the density laws from the nearby thin disk to the 
distant halo. 
We also have the range in latitude and longitude to break the apparent normalization-scale height degeneracy problem 
described in Paper I.

\subsection{Observations}

We observed twelve fields over the course of nine observing runs from August 1993 through September 1997 with the Las 
Campanas Swope 1-meter telescope.  These observing runs spanned a total of 51 nights, 31 of which were 
photometric.  From August 1993 through January 1995, we used the $2048^2$ Tektronix 3 CCD camera at Cassegrain 
focus with a pixel scale of $0\farcs61$ per pixel.
In September of 1995, we switched to the larger pixeled ($0\farcs69$ per pixel) and more sensitive Tektronix 5 CCD chip.  The last observing 
run used the SITe 1 CCD chip, a near twin of the Tektronix 5.
While data from all three chips have been 
reduced through the pipeline described in \S2-4, the subsample used for this particular study is
entirely from the Tek 5 and SITe 1 chips.

Each Selected Area was divided into a grid of subfields.  These grids were designed to provide maximum spatial coverage of the Selected
Area while allowing a small overlap with adjacent subfields for photometric comparison or bootstrap calibration of non-photometric observations.
The fields were first observed in Johnson $BV$ and Thuan-Gunn $i$ filters.  After the switch to Tek 5, the 
Cousins $R$ filter was added to the program.  Both the $B$ and $R$ filters were of the Harris
prescription.  A number of subfields were also observed with the $CuSO_4$ or 
Harris $U$ band filter.  All of
these passbands were calibrated to the Landolt (1992) system (see \S3.2) which uses Johnson $UBV$ and Cousins $RI$
filters.  Color terms in the transformations (see below) removed discrepancies between our observational photometry
system and Landolt's standard system.

Each subfield in the grid was observed in a sequence of $BVI$, $RI$, $BVRI$ or $UB$ exposures.  Long and short exposures were obtained
in each filter to increase the dynamic range of the sample.  Short exposures were 15-40 seconds, depending on observing
conditions and filter.  Long exposures were typically 450-1200 seconds.  This provided photometric coverage to an
estimated average depth of $(B,V,R,I)=(21.1,21.4,21.5,20.6)$ with stars as faint as 
$(B,V,R,I) = (23.0,22.6,22.2,21.5)$ photometered in our best frames.  The $U$ band observations proved particularly difficult
and exposure times as great as 1800 seconds were sometimes required to reach our target photometric
depth.  These frames have not yet been thoroughly evaluated to test their photometric depth but 
preliminary indications are that they are complete to $U\sim21$.
Tek 5 $BVI$ observations typically reach a magnitude fainter than the Tek 3 $BVI$ observations.

In July of 1996, we observed two additional fields to constrain asymmetry
across the $l=0^{\circ}$ meridian.  These fields were similar in coordinates to SA184 and SA107
but reflected across the Galactic meridian (and thus these ``anti-fields" were given the labels ASA184 and ASA107).
In July of 1997, we observed subfields three and eight of ASA184 with the 
2.5 meter du Pont telescope at Las Campanas in an attempt to determine if those 
fields may be contaminated by the tidal stream of the Sagittarius dSph 
(described in Paper III).  The 2.5 meter telescope, in addition to having larger aperture, boasts
a finer plate scale than Swope.  Thus, a side benefit of the exploration was the provision
of two deep exposures with which to compare our program observations
and to evaluate directly our completeness level and classification success (\S4.3).

\section{Photometric Pipeline}

\subsection{Reductions}

The CCD images were trimmed and corrected for bias, overscan, flat field and illumination effects with the 
CCDRED package in IRAF.\footnote{
IRAF is distributed by the National Optical Astronomy Observatories,
which are operated by the Association of Universities for Research
in Astronomy, Inc., under cooperative agreement with the National
Science Foundation.}  We found that the best illumination correction was produced by using the DIMSUM package to mask 
bright stars in our deeper program exposures and then using these masked images as additional sky images 
to combine into sky flats.
We used the IMREPLACE process of IRAF to mask bad pixels to typical sky values.

We performed photometry on the data using the IRAF version of the 
DAOPHOT photometry package (Stetson 1987).  
A Moffat profile was used with quadratic geometric variation.  The Swope data 
proved to be an excellent source of wide field photometry.
Typical DAOPHOT $\chi$ values of the PSF
fit were 0.01-0.03, with few frames worse than 0.04.  Visual inspection of the frames revealed
star subtraction to be excellent.

The fields observed in our program are uncrowded and one could argue that simple aperture 
photometry would be adequate.  We have found, however, that the structural parameters produced by DAOPHOT are 
useful for object classification (see \S4).  We have also found that aperture photometry is noticeably inferior to PSF photometry at 
the faint end of the data.  Finally, PSF photometry can interpolate around chip edges and bad pixels 
(a significant problem with the Tek 3 chip) to render a more complete survey.

\subsection{Frame Matching, Calibration and Photometric Consistency}

During our observing runs, we periodically observed Landolt (1992) standard stars with a broad color
range over a large range of airmasses.  In addition, more
than half of our survey areas contain Landolt standard stars.
The raw photometry was calibrated onto the standard system through several steps.  
We first derived transformation constants for each observing run, using the formulation:
	
\begin{center}
	$U = m_u + u1 + i2 \times X_u + u3 \times (U-B) + u4 \times (U-B) \times X_u + u5 \times (U-B)^2$
	$B = m_b + b1 + b2 \times X_b + b3 \times (B-V) + b4 \times (B-V) \times X_b + b5 \times (B-V)^2$
	$V = m_v + v1 + v2 \times X_v + v3 \times (B-V) + v4 \times (B-V) \times X_v + v5 \times (B-V)^2$
	$R = m_r + r1 + r2 \times X_r + r3 \times (R-I) + r4 \times (R-I) \times X_r + r5 \times (R-I)^2$
	$I = m_i + i1 + i2 \times X_i + i3 \times (R-I) + i4 \times (R-I) \times X_i + i5 \times (R-I)^2$
\end{center}
 
\noindent where, in the visual passband, $V$ is the apparent magnitude, $m_v$ is the instrumental magnitude, $B-V$ is the 
color and $X_v$ is the airmass of the observation; corresponding terms apply to the other four passbands.
These equations were solved via the matrix inversion technique described by 
Harris (1981) through a program that allows the user to fit one of the above equations 
interactively.  For all five filters, the color-squared and color-airmass terms were found to be 
insignificant and were
subsequently dropped from the solution with no loss of precision or accuracy.  Solutions converged at 
better than 0.03 magnitudes RMS precision in all five passbands for all photometric nights.  Most 
solutions were better than 0.01 magnitudes.

Our ultimate goal was for each Selected Area photometry catalogue to be mosaiced from
photometrically interlocked subfields and calibrated to the Landolt system.  The first step
was to combine the multifilter data for each subfield into a calibrated photometry
catalogue.  The multifilter data were matched using the DAOMASTER code included with DAOPHOT.  These 
instrumental subfield catalogues were 
then transformed to the standard system using the equations above and a program designed to resolve
any zero point differences between multiple observations in the same passband.  The photometry from 
each image was compared to the average of all photometric images in its passband.  The resulting zero 
point difference was then added to the $v1$ term.  This process was performed iteratively until the 
frame-to-frame residuals were reduced below a user-specified level (usually .001 
magnitudes) in each passband.  Non-photometric observations were included in the solution and were
correctly transformed to the average photometry of the photometric frames (see below).  Once zero point
differences were removed, the magnitudes were averaged to produce the final measure.  Typical
zero point difference were a few hundredths of a magnitude.

The next step was to interlock the photometry by comparing overlapping photometric subfields.  We
found systematic zero point residuals of 0.01-0.1 magnitudes, possibly resulting from changes in sky 
transparency or variations in the extinction coefficients.

A number of subfields had only non-photometric observations.  To calibrate these subfields, 
we used the stars at the edge of the images that overlapped
photometric subfields for bootstrap calibration.  Direct comparison allowed the calculation of a 
separate zero point term for each CCD frame.  Color terms were then set to the values derived for the 
observing run on which the non-photometric frame was taken.  The bootstrap calibrations showed no sign of non-linearity 
in magnitude or color.

Once these residuals were corrected for, we identified Landolt standard stars within each Selected Area
and compared our final photometry with the Landolt measures to identify any remaining zero point
problems.  The residuals of these comparisons are within the expected photometric scatter across the 
magnitude range (see Figure 3) and the average residual of standard stars within any particular 
Selected Area is within the stated photometric error.

Astrometry for our stars was derived from the USNO SA2.0 catalogue (Monet et al. 1986) using the STSDAS TFINDER
package.  The subfield catalogues were then combined using this astrometry to match the stars.
The $RI$ data utilized for the photometric parallax program discussed below do not have multiply observed stars removed (i.e.,
stars at the edge of each CCD chip, about 25\% of the sample).  To account for this effect upon our 
analysis, we have calculated the solid angles below as though the subfields did not overlap.  In a 
statistical survey, the effect on the analysis is negligible because multiply-represented stars appear
only as a function of their location within each mosaiced SA field.
Future contributions will be from 
a catalogue with the multiples averaged together.

\section{Completeness and Classification}
	
The analysis of deep starcounts can be critically affected by faint galaxy contamination
(see Paper I for discussion).  At the magnitude limit of our data ($R \sim 21$), galaxies outnumber stars
by an order of magnitude.  Because extragalactic objects are distributed uniformly over the sky and are 
predominantly faint, failure to fully remove this source of contamination will, for example, skew 
evaluations of the distant halo density distribution toward high axial ratios 
(i.e., a rounder halo) and shallow density gradients.

As shown in Paper II, photometric information alone is rather ineffective at discriminating stars from galaxies.
Approximately two thirds of galaxian objects lie within 2 $\sigma$ of the stellar locus in multicolor space.  
This is hardly surprisingly since galaxies are, after all, comprised of stars.
Morphological information has proven to be the single best discriminant and Paper II identified 
four morphological classification methods -- ellipticity, the $\chi$ parameter of DAOPHOT, and 
two measures of image compactness -- with which to seperate stars from galaxies.

Those methods, though effective, were designed for a small set of HST and Keck observations and 
may not be as effective for data taken at a much coarser pixel scale ($0\farcs61-0\farcs69$ per pixel).  Moreover, the extensive
data set produced by our program demands more automated methods.  The DAOPHOT program produces
enough information to provide reasonable object classification.  
The DAOFIND algorithm uses estimates of roundness and sharpness to filter the initial list of 
potential stars.  Additionally, ALLSTAR produces measures of $\chi$ and SHARP.  We have used
these measured to construct a simple but effective classification engine.

Figure 4 shows the morphological parameters of objects in a typical 
field as a function of magnitude.  Brighter than a certain magnitude, there are clearly two loci of objects 
corresponding to stars and galaxies.  The magnitude at which the loci merge is the faintest magnitude at which we 
can confidently discriminate stars from galaxies.  All objects fainter than this classification limit
must be considered potentially misclassified.

Since our analysis is limited to objects brighter than the classification limit, the classification limit
replaced the magnitude limit in our evaluation of survey completeness.  
Of course, the classification limit and the magnitude limit of any 
particular CCD frame should be correlated since the factors that limit successful image
classification - seeing, sky brightness, pixel scale - are also those which affect photometric depth.

Our technique for star/galaxy separation was to inspect the magnitude-$\chi$ and magnitude-SHARP 
distribution of each subfield to identify the classification limit.  Objects below this limiting magnitude
were selected as stars if they had $\chi<2$ and $-0.4 < SHARP < S_u$, where $S_u$ is the value of SHARP
where the galaxian and stellar loci merge\footnote{Although a more stringent $\chi$ limit would be
reasonable, we found that some bright stars had elevated $\chi$ values, possibly a result of DAOPHOT 
underestimating the expected error or nonlinearity near the saturation limit of the CCD.  We therefore 
used a more generous upper limit.  The galaxies allowed by this generous $\chi$ limit at the faint end are 
removed by the more stringent SHARP limit.}.  $S_u$ varied from image to image
depending on observing conditions but was typically 0.1-0.15.   DAOFIND 
parameters were left at the IRAF defaults of $-1<ROUND<1$ and $0.2<SHARP<1.0$.
This technique was the product of extensive testing of the data pipeline using
the methods detailed below.

\subsection{ADDSTAR}

Our first pipeline test used the DAOPHOT task ADDSTAR to implant a set of 1000 artificial
stars, evenly distributed between $V=15$ and $V=25$, to eight sets of Tek 3 and Tek 5 observations of 
varying quality.  This program adds stellar images to the CCD frame by scaling the PSF to the
appropriate magnitude and adding realistic noise.  The revised CCD images images were photometered and 
classified, after which the artificial stars were extracted.
Figure 5a shows the combined result of this test before object classification.  We
recover 100\% of the stars to and beyond the classification limit of each frame.  The recovered fraction
reaches the 50\% level approximately 1.5 magnitudes fainter than the classification limit and the 
0\% level 3 magnitudes below the limit.  This result is independent of the actual value of the 
imaging limit and consistent from the worst data (bright sky background, $> 2\farcs5$ 
seeing) to the best data (low sky background, $1\farcs0$ seeing).
Figure 5b shows the result of a high resolution version of this test, in which we added 
1000 stars in a 3-magnitude wide bin centered on the classification limit.  We find that the recovered 
fraction is still 100\% at the classification limit, dropping off to the  
50\% level 1.2 magnitudes fainter.

We then ran the extracted measures through the classification pipeline to determine how many artificial
stars were misclassified as galaxies.  Figure 6 shows retained fraction as a function of magnitude with respect to
the classification limit.  We have found that our method leaves the artificial stars 96-100\% intact.  The 
percentage of stars misclassified appears to be independent of magnitude above the classification
limit.  Even the high resolution data show, 
at most, a drop from 99\% to 96\% in the recovered fraction to the classification limit.

The measured photometry of our artificial stars has unrealistically small errors, producing unrealistically 
low $\chi$'s.  This results from the analytical PSF being only an approximation to the light distribution
of real stars, but is a perfect description of artificial stars.  Figure 7 shows the 
distribution of the artificial stars in magnitude-$\chi$ and magnitude-SHARP space, which can be 
contrasted with Figure 4.  Our evaluation of the efficacy of our pipeline is uncompromised
by the low $\chi$ values because the distribution of SHARP values, the primary classification discriminant, 
is similar for both the artificial and real stars.

\subsection{ARTDATA}
	
Our second pipeline test used the ARTDATA package in IRAF to create a 
sample of synthetic starcount observations.  We generated a catalogue of stars distributed according
to a prescribed Galactic structure model (the interim model of Paper I), then added 700 extragalactic 
objects to the sample, with a power law distribution of $N(A) \propto 10^{0.18*m}$
This distribution is slightly shallower than galaxy distributions 
derived from deep $R$ band galaxy count surveys (e.g., Crawford et al. 1999)
and produces a fair level of contamination near the classification limit of 
the artificial data.  Two types of galaxies were added - a de Vaucouleurs spheroid and an 
exponential disk - in a variety of orientations.

This catalogue was used to add objects to a set of synthetic $BVRI$ images, generating realistic
Poisson noise and adding a typical sky background.  These ``observations" were 
then evaluated through the photometry and classification pipeline.

Figure 8a shows the recovered fraction as a function of magnitude for the artificial data
before object classification for all images in all passbands.  The overall trend is similar to the ADDSTAR test.  Figure 8b shows the 
recovered fraction of stars after classification.  The recovered fraction drops near the classification
limit.  Exploration of the data has shown this drop to be a result of the binning.
Our methods retain 92\% of the stars to the classification limit.
	
Figure 9 shows the magnitude-$\chi$ and magnitude-sharp distributions of 
the artificial galaxies measured in our data.  607 galaxies were 
photometered in at least two frames and 110 of those galaxies were above the classification
limit.  The classification parameters eliminated all but two of those galaxies - both 
faint de Vaucouleurs galaxies.

We evaluated the preselection parameters (sharpness and roundness) as classification
tools by broadening the limits of DAOFIND.  Changing the sharp limit added a handful of objects
to the sample, while expanding the roundness limits (from -1:1 to -2:2) added 176 
detections to the initial detection sample of the nearly 3000 objects.  Visual inspection revealed
the new detections to be mostly noise spikes and a few very bright stars.  All but three bright
stars among the new detections were rejected by the pipeline.  From this, we conclude that the 
roundness and sharpness parameters of DAOFIND primarily separate out cosmic rays.  The preselection 
parameters improperly remove less than 0.1\% of the stars in observation and the stars removed 
are entirely bright saturated stars.

\subsection{Deep ASA184 exposures}

Our final test compared the two deep ASA184 exposures used for our search for the Sagittarius stream 
(Paper III) against overlapping Swope exposures initially intended to constrain halo 
asymmetry.  We identified 430 objects observed in both the Swope and du Pont images that
were brighter than the Swope-based classification limits for the two subfields.  We then classified
these objects using the finer resolution of the du Pont data.
The distribution in magnitude-$\chi$ and magnitude-SHARP for the matched objects is
shown in Figure 10.  Of 36 galaxies photometered on both data sets, the pipeline correctly 
classified all of them on the Swope images.  Of 394 stars in common, all but ten were correctly 
classified on the Swope images.  Visual inspection revealed that the misclassified stars were near chip 
edges, bad columns or bright saturated stars.  Figure 11 shows the recovered fraction as a function
of magnitude relative to the classification limit of the Swope data.  The recovered fraction of stars 
is 90-100\% at this limit and the 50\% level is reached at 0.9 magnitudes below the limit.  This is 
similar to the results of the artificial star tests.

\subsection{QSO Contamination}

The classification methods discussed above are robust for galaxies that are morphologically
distinct from stars.  QSOs, however, are pointlike enough to be easily confused with stars.  
In Paper I, we noted
that those objects can contribute over 25\% of the blue ``starcounts" at faint magnitudes.

QSO contamination is less of a problem in our study because these extragalactic objects are 
predominantly blue ($B-V < 0.6$, $R-I < 0.4$).  Conveniently, the stars
we study here are redder than this range (see \S5.2).  Nevertheless, some 
contamination may remain and the only way to mitigate the effect is to make a statistical correction to 
the counts.

To estimate just how much of a contribution QSOs will make to our starcounts, we estimated
the number of these extragalactic objects that would fall into the faintest ($20<R<21$), bluest ($0.4<R-I<0.6$) 
region of the color-magnitude area from which photometric parallaxes are derived.  Using the quasar 
luminosity function of Kron et al. (1991), convolved with the SDSS color distribution of the 
quasars listed in Richards et al. (2001), transformed from SDSS to standard filters using the 
transformations of Fukugita (1996), we estimate that this color-magnitude bin should have a 
contamination level of 32 galaxies per square degree.

Is this significant?  This color-magnitude bin contains 150-500 
stars per square degree, depending on Galactic latitude.  This would mean that QSOs inflate
the starcounts by 5-25\%, with the starcount inflation greatest at high Galactic latitude.  We repeated
part of our analysis (\S6) with the counts statistically corrected for compact galaxy contamination and 
found only a small effect on the derived density laws.  We have thus chosen to leave this correction 
out of our analysis.

\subsection{Extragalactic Contamination - Overview}

Considering the results of the tests of our pipeline, we are confident 
that we have photometered all stars to the stated imaging limits 
of each subfield, successfully stripped out nearly 100\% of the 
galaxies from the starcounts to the classification limit and correctly classified 95-100\% 
of the stars down to the classification limit.

Even more critical than the fraction of stars lost either through incompleteness or 
object classification are potential magnitude- or
color-dependent biases induced.
Our results show that the fraction of stars retained by the pipeline is a constant 
as a function of magnitude to the classification limit.

\section{Photometric Parallax}

There are many experiments that can be performed with a sample of starcount data.  Our ultimate goal is 
to expand on the work in Paper I, generating a complex Galactic structure model and making comparisons 
between synthetic data generated by that model and real observed starcounts from our more extensive CCD
data.  In this article, we make a first evaluation of the density laws of the 
Milky Way stellar populations using a more 
simple method.  We take a subsample of our data, $RI$ photometry from seven of the Selected Areas, and 
use photometric parallax distance estimates of these stars to constrain the density laws of the thin 
disk, thick disk and 
halo.  The $RI$ data were chosen because of the near uniformity of the data and the relative insensitivity to 
reddening of the $RI$ passbands.  
Future contributions will exploit larger portions of the $UBVRI$ data in an effort to constrain both 
the luminosity functions and spatial distributions of the Galactic stellar populations.

\subsection{Dereddening}

The photometry was dereddened using the high resolution COBE/DIRBE maps
of Schlegel et al. (1998), which closely match the earlier work of 
Burstein \& Heiles (1982).  The reddening for the fields included in the photometric parallax study was 
generally small, with most of the fields near $E_{B-V} \leq 0.05$ and only SA107 reddened by as much as 
$E_{B-V} = 0.1$.  Some program fields not included in this analysis had higher values and SA95 showed 
evidence of differential reddening across the field.  Interstellar 
reddening was corrected on a star-by-star bias by interpolating the Schlegel et al. maps at the position 
of each star and using the reddening coefficient for the Landolt $UBVRI$ system
derived in Schlegel et al.

The blue edge of the field photometry (see Figure 12) reflects the color of the main sequence turnoff
stars at any particular magnitude.  Previous studies (see, e.g., Unavane et al. 1996, C01) have shown the color of the blue 
edge as a function of magnitude to be remarkably consistent over a large range of $(l,b)$.  It 
should thus serve as a reasonable landmark with which to check the derived reddening.  Our subsequent 
study of the blue edge (Siegel et al., in preparation, hereafter Paper V) has revealed this blue
edge color to 
align well in all of our program fields, which indicates we have made an accurate correction.  The
blue edge of SA95, a field which has patchy foreground reddening, was changed from an indistinct blur into 
a tight line after reddening correction (see Figure 12 and Paper V).

In the present analysis, we have assumed that dust obscuration is entirely in the foreground of our
stars, which is inappropriate for the stars nearest the plane.  Nevertheless, this assumption should be 
valid for the vast majority of stars because the scale height of the reddening layer is approximately 100 pc 
(Chen et al. 1999), which is far smaller than the distance of the bulk of the old disk, thick disk and 
halo stars we study.  Even with a complex extinction model, the bulk of our stars would have nearly the full 
reddening correction applied.

\subsection{The $M_R(R-I)$ Relation}

The first step in determining photometric parallaxes is estimating the absolute
magnitude of the program stars.  We have derived a color-absolute magnitude relation in
the $RI$ passbands for dwarf stars and
assumed {\it ab initio} that the bulk of our stars are faint dwarfs.  The color-absolute magnitude
relation utilizes parallaxes from the ESA Hipparcos
catalogue of nearby dwarf
stars and corresponding photometry from Bessell (1990) and Leggett (1992).\footnote{Available online at 
http://dep.physics.upenn.edu/$\sim$inr/cmd.html}  The catalogue was 
cleaned of known binaries, stars with poor parallaxes ($\sigma_{\pi}/\pi > 0.2$) and stars clearly 
removed from the trend of main sequence dwarf stars.  After the application of the Lutz-Kelker (1973) 
correction, we fit a two-part relation of:

\begin{eqnarray}
M_R = -6.862 + 61.375 (R-I) - 108.875 (R-I)^2 + & \nonumber\\
90.198 (R-I)^3 - 27.468 (R-I)^4 & 0.4 \leq R-I < 1.0\\
M_R = -114.355 + 408.842 (R-I) - 513.008 (R-I)^2 + & \nonumber\\
286.537 (R-I)^3 - 59.548 (R-I)^4 & 1.0 \leq R-I < 1.5
\end{eqnarray}

\noindent based on samples 230 and 195 stars, respectively.  The fit is shown in Figure 13.  These 
relations have a small discontinuity of approximately 0.1 magnitude at $R-I = 1.0$.  The cutoff
at $R-I$=0.4 eliminates potential uncertainties in the photometric parallax due to the main sequence
turnoff as this is well-redward of the MSTO of even old populations.

The color-absolute magnitude relations have an uncertainty of 0.2-0.3 magnitudes.  Therefore, the distance
to any particular star has a moderate degree of error.  If this scatter created corresponding 
{\it random} distance errors, it would be of small concern in a statistical analysis of 70,000 stars. 
However, the large absolute magnitude uncertainty plays a significant role in creating a 
{\it systematic} error in the distance estimates (Malmquist Bias) that will mask itself as a spatial 
trend.  We now discuss Malmquist Bias and three other expected systematic biases -- subgiant 
contamination, subdwarf bias and binarism -- that must be accounted for.

\subsection{Malmquist Bias Correction}

In a conical magnitude-limited volume, the distance to which 
intrinsically bright stars are visible is larger than the distance to which
intrinsically faint stars are visible.  The effect of this is that
brighter stars are statistically over-represented, and the derived 
absolute magnitude are too faint.  This effect, 
known as Malmquist Bias (1920), was formalized into the
general formula (shown here for our survey magnitudes and colors):

\begin{equation}
M(R)=M_{0} - \frac{\sigma^2}{0.4343} \frac{dlog A(R)}{dR}
\end{equation}

\noindent with the error in the magnitudes given as:

\begin{equation}
\sigma^2 = \frac{dM(R-I)}{d(R-I)}^2 * \sigma_{R-I}^2 + \sigma_{M_R}^2
\end{equation}

\noindent where $M_{0}$ is the absolute magnitude calculated for any star's $R-I$ from the $M_R=f(R-I)$ relation, 
$A(R)$ is the differential counts evaluated at the apparent magnitude $R$ of any star,
and the errors correspond to the effect of photometric error propagated through the $M_R=f(R-I)$ 
relation as well as the intrinsic astrophysical scatter in that relation, respectively.

A sophisticated computer model that produces artificial comparison data could be made to incorporate Malmquist
bias naturally.  Our simpler analysis in the distance-density domain requires a
correction to the absolute magnitudes of the stars.  It should be noted that
this absolute magnitude correction is only applied in {\it statistical evaluations}
of the stars.  For any star considered as an individual object, this correction
is inappropriate.  When the stars are considered {\it en masse} however, each absolute magnitude
must be corrected.  This is analogous to the Lutz-Kelker correction applied to correct a 
$\sigma_{\pi}/\pi$-limited sample bias in our
derivation of the color-absolute magnitude relationship and subdwarf bias (\S5.2 and 5.5).

To make the Malmquist corrections, we produced a Hess diagram
of each field, binned in 0.1 color intervals and 0.5 magnitude intervals.  Faint magnitude
bins were corrected to reflect the declining number of subfields complete to the corresponding
depth.  For each binned color range, the $A(R)$ function was fit with a power
series.  The parameters of the fit change smoothly
with color, which indicates a consistent fit to the color-magnitude-counts 
surface.  For each star, the derivative for its color bin was evaluated at its magnitude and
combined with the observational color error and the intrinsic scatter in the $M_R=f(R-I)$ relation 
to derive the correction to its absolute magnitude.

The most distant stars in our sample are only marginally affected by Malmquist bias since the differential 
counts level out at faint magnitudes (as expected for an $R^{-3}$ power law density
distribution that approximates the Galactic halo).  This causes the first derivative of the counts to 
be near zero, resulting in a minimal Malmquist correction.  Nearby stars have a sharper gradient 
in differential counts, resulting in more substantial corrections.

\subsection{Subgiant Contamination}

Our photometric parallax method assumes simplistically that every star in the data set is an unevolved 
dwarf star.  While evolved giant and subgiant stars are fewer than main sequence stars by an 
order magnitude, giants and subgiants are much brighter than their main sequence brethren and can therefore 
be detected to greater distances and larger volumes.  Subgiants could therefore contribute to the 
counts in a ratio much greater than predicted from the luminosity function.

Subgiant contamination was once speculated to be the source of the thick
disk signal detected by GR83 (Bahcall \& Soneira 1984, in which the subgiants are referred to as ``giants").  
This argument has since been disproven 
(see Paper I), but the role of subgiant (and, for surveys at brighter apparent magnitudes, giant) 
contamination is a concern for any use of photometric parallaxes based on dwarf color-magnitude
relations.  The problem can be addressed on a statistical basis with an analytical correction.  

At any particular color (and corresponding dwarf absolute magnitude $M_D$) the dwarfs at
any apparent magnitude $m$ are at a distance $r_D = 10^{ (m-M_D+5)/5}$.  At this
magnitude, the identically colored subgiants/giants of absolute magnitude $M_{SG}$ are a distance $r_{SG} = 
10^{ (m-M_{SG}+5)/5}$.  The number of giant contaminants, $N_{SG}$ in the magnitude interval 
($m-\delta_m,m+\delta_m$) and corresponding distance interval 
($r_{SG}-\delta_{r_{SG}},r_{SG}+\delta_{r_{SG}}$) in solid angle $\Omega$ can then be defined as:

\begin{equation}
N_{SG} = \Omega \kappa_{SG,D} \int_{r_{SG}-\delta_{r_{SG}}}^{r_{SG}+\delta_{r_{SG}}} \rho(r_{SG}) r_{SG}^2 dr_{SG}
\end{equation}

\noindent where $\kappa_{SG,D}$ is the intrinsic ratio of subgiants to dwarfs in a stellar population 
(as determined from the luminosity function) and $\rho$ is the density of dwarfs as a function of 
distance.  This is essentially the von Seeliger equation evaluated for a single spectral type.  In the 
limit where the density law does is relatively constant over the distance interval $\delta_{r_{SG}}$ 
and $\delta_{r_{SG}} << r_{SG}$, this can be rewritten as:

\begin{equation}
N_{SG} = \Omega \kappa_{SG,D} \rho(r_{SG}) \times 2r_{SG}^2\delta_{r_{SG}}
\end{equation}

The corresponding dwarf stars are evaluated in the same magnitude interval ($m-\delta_m,m+\delta_m$).  Our
starcount analysis is directed at evaluating the number of dwarf stars in the
distance interval ($r_D-\delta_{r_D},r_D+\delta_{r_D}$).  We therefore derive the 
density of dwarfs in this magnitude interval over the volume element 
$V_D=2 \Omega r_D^2 \delta_{r_D}$.  The evolved stars
contribute $N_{SG}$ extra counts to this volume interval, producing a density inflation of:

\begin{equation}
\rho_{SG} = N_{SG}/ V_D = \kappa_{SG,D} (r_{SG}^2 \delta_{r_{SG}}) \rho(r_{SG}) / (r_D^2 \delta_{r_D})
\end{equation}

For any color or dwarf absolute  magnitude, we can explicitly define
the subgiant distance $r_{SG}$ in terms of the difference in absolute magnitudes between the dwarfs 
($M_D$) and subgiants ($M_{SG}$) or the ratio of luminosities $\frac{L_{SG}}{L_D}$:

\begin{equation}
r_{SG} = r_D * 10^{(M_D - M_{SG})/5} = r_D * \sqrt{\frac{L_{SG}}{L_D}}
\end{equation}

\begin{equation}
\delta_{r_{SG}} \sim \delta_{r_D} * \sqrt{\frac{L_{SG}}{L_D}}\\
\end{equation}

Substituting equations 10 and 11 into equation 9 yields:

\begin{equation}
\rho_{SG} =  \left (\frac{L_{SG}}{L_D}\right )^{1.5} \times \kappa_{SG,D} \times \rho \left (r_{D} * \sqrt{\frac{L_{SG}}{L_D}} \right )
\end{equation}

At a constant $\kappa_{SG,D}$ and ($\frac{L_{SG}}{L_D})$, the gain in accessible 
subgiant detection volume due to greater subgiant luminosity is matched by the loss due to declining 
density if the density falls as $r^{-3}$ or steeper.
For most of our fields, the distance vector is close to the Galactic radial vector at large
distances and the density should fall off as $r^{-3}$ in an $R^{-3}$ power law or de Vaucouleurs
profile halo.  
SA107/SA184, however, are an exception.  Because these two fields point into the Galaxy, models with
spherical or near spherical halos will predict a {\it rise} in density between 5 and 10 kpc.  Thus, 
evolved stars could be a more significant contaminant in those fields.  The bias must be quantified
and corrected if it is significant.

$\frac{L_{SG}}{L_D}$ and $\kappa_{SG,D}$ are a function of color.  For very red stars, $\kappa_{SG,D}$ 
is low enough and $\frac{L_{SG}}{L_D}$ high enough that evolved star contamination will be minimal. For 
late G-/early K-type stars, 
subgiant contamination might be more significant.  We evaluated the effect on our density law 
derivation for our brightest bluest dwarf stars ($5.8 \leq M_R < 6.8$).  We
determined $\frac{L_SG}{L_D}$ and $\kappa_{SG,D}$ with an [Fe/H]=-1.5, 16 Gyr isochrone from Bergbusch \& 
Vandenberg (2001, hereafter BV01) for the distant halo and an [Fe/H]=-0.6, 12 Gyr isochrone to represent the nearby 
thick disk and old disk.  The resultant values ($\frac{L_{SG}}{L_D}\sim26, \kappa_{SG,D}\sim.03$) 
were put into equation 11 to correct the predicted density for any particular model as a function of 
distance by:

\begin{equation}
\rho_{SG}= 4 \rho(5r_D)
\end{equation}

The total number of stars observed in a distance bin corresponding to a dwarf star apparent magnitude
range of ($m-\delta_m,m+\delta_m)$ is the sum of dwarfs and contaminating evolved stars:

\begin{equation}
\rho_{total} = \rho(D) + \rho (SG)
\end{equation}
\begin{equation}
\rho_{total} = \rho(r_D) + 4 \rho(5r_D)
\end{equation}

This is a worst case scenario because the contamination levels drop off sharply for stars with fainter absolute
magnitudes.  We have found no significant effect of subgiants upon our analysis, even in SA184/SA107.  The worst
case scenario resulted in no significant change in the derived parameters of our best fit starcount
models (\S7) but improved the $\chi^2$ of the fit from 2.7 to 2.2.  The reason for this becomes obvious in light
of the flattened halo models favored by our analysis in \S7.  Such models reduce the stellar density at the 
distance of potential subgiant contaminants compared with spherical halo models.

Our conclusion is that, even for our bluest and brightest main sequence stars, subgiant contamination has minimal 
impact on our density law derivation.  We have therefore elected to exclude this contribution from the 
density law derivation rather than attempt a more complex solution to account for the sharp slope of 
$\frac{L_{SG}}{L_D}$ and $\kappa_{SG,D}$ with absolute magnitude and color.

\subsection{The Subdwarf Correction}

The solar $M_R=f(R-I)$ relation we define above is based on nearby metal-rich stars.  However, this
relation is systematically discrepant from the relation(s) for metal-weak stars due
to the effect of line blanketing.  Metal-rich stars are generally redder and fainter than their 
metal-weak, equal mass counterparts due to such effects.  While the effect of line blanketing is
small for G-type stars in the $R$ passband, late-type K 
and M dwarfs develop strong molecular bands in the $R$ passband with increased abundance, particularly from TiO and 
VO.  The net result is that the main sequence of metal-poor stars is blueward of
the solar-metallicity sequence.  If photometric parallaxes are derived
from stellar colors, the distance to metal-poor stars will thus be systematically overestimated.
Because the mean abundance of stars changes with spatial position, ignoring the metallicity effects in
the derivation of photometric parallax results 
in systematic errors for the derived spatial structure.

Deriving a metal-poor $M_R=f(R-I)$ relation is problematical since few metal-poor stars have both precise trigonometric
parallaxes and high-quality $RI$ photometry.  We have adopted a strategy similar to that of Gizis \& Reid 
(1999).  The relation given in \S5.2 is for local stars that have an average metallicity of approximately [Fe/H]=-0.2.  
We have defined a second relation for the metal poor ``sd" stars of Gizis (1997), which have an average 
estimated mean [Fe/H] of -1.2.  The relation for these stars, with Lutz-Kelker correction, is:

\begin{equation}
M_R = 2.03 + 10.0 \times (R-I) - 2.21 \times (R-I)^2
\end{equation}

\noindent We find that this quadratic equation is consistent with the two-part linear calibration of Gizis \& Reid (1999) and consistent
with the bright subdwarf tail of theoretical isochrones from BV01 (see Figure 14).  For a star of given
metallicity and color, its absolute magnitude can be estimated by linear interpolation between the two
derived ridgelines and linear extrapolation beyond the metal-poor ridgeline.\footnote{Although the 
[Fe/H]=-2.0 stars form an apparent sequence fainter than the [Fe/H]=-1.2 ridgeline, we have elected not 
to establish a third ridgeline.  Our reasons are that the uncertainties in the [Fe/H]=-2.0 subdwarfs are 
large, there are no [Fe/H]=-2.0 stars bluer than $R-I=0.75$ with which to extend the sequence, and
we have truncated the metallicity gradient so that no star is assumed to be more metal-poor than [Fe/H]=-1.5.}

Following GR83, we assign metallicities to stars based on their location in the Galaxy, using an 
empirical metallicity
distribution.  We assume that derived mean metallicities for tracer stars (e.g., K giants) apply to 
the dwarf population and modify the vertical abundance gradient of Yoss et. al. (1987) to:

\begin{eqnarray}
0< z < 0.7 kpc & [Fe/H]=-0.4 \times z \nonumber\\
0.7 < z < 7.5 kpc & [Fe/H]=-0.28 - 0.18*(z-0.7 kpc)\\
7.5 kpc < z & [Fe/H] = -1.5 \nonumber
\end{eqnarray}

\noindent where $z$ is in units of kpc.
The first two abundance gradients are directly from Yoss.  But, unlike Yoss et al., we truncate the gradient at 
the point it reaches [Fe/H] = -1.5, a likely average metallicity value of the halo (Carney et al. 1996).

A similar 
empirical abundance gradient is derived by Trefzger et al. (1995).
This model does not account for any radial variation in metallicity.  The existence of such a
radial gradient in the field stars of any population beyond the thin disk has yet to be demonstrated 
(see, e.g., Rong et al. 2001).
While vertical metallicity gradients have been refuted for the halo (Searle \& Zinn, 1978; Carney et al. 1990) 
and are debatable for the thick disk (see, e.g., Rong et al. 2001; cf. Gilmore et al. 
1995, R96), such analyses have attempted to isolate each population and assess their abundance distributions
independently.  This is a risky venture because of the difficulty in separating overlapping stellar populations
and because apparent metallicity gradients may be the product of
transitions from monometallic regions dominated by metal-rich populations to monometallic regions 
dominated by metal-poor populations.
However, one can make a comparison between our modified abundance gradient and one produced
by a favored model abundance distribution.  Adopting the density laws derived in \S7 and 
assuming traditional average abundances of [Fe/H]=0.0, -0.6 and -1.5 for the thin thick, thick disk and halo, 
respectively, we find the Yoss gradient is slightly ($\sim0.1-0.2$ dex) too metal-rich 
at Galactic heights dominated by the thick disk (2-6 kpc).  A slope of 0.21 
dex/kpc in the second relation produces a better match to the model abundance
gradient but the difference in photometric parallaxes is negligible.
Given the similarity in results, we opt for simplicity and retain the
empirical abundance gradient, which also has the advantage of avoiding a
priori  assumptions concerning population mixture ratios.

With each star, a preliminary distance is determined using the near-solar metallicity relation.  We then 
derive a metallicity based upon the star's height above the plane and linearly interpolate from the two 
color-magnitude relations to derive a corrected absolute magnitude.  This process is performed 
iteratively until the distance converges to within 5 parsecs.  The effect of this correction upon the
derived density laws is a subtle steepening of the density gradient for stars at large heights above
the Galactic midplane.

We note that systematic errors in our subdwarf correction would result in systematic errors in our derived
density laws.  For example, systematically underestimating the abundance of distant stars would cause us
to underestimate the absolute magnitudes and underestimate the distance, resulting in steeper density
gradients and reduced scale heights.  Such concerns are addressed more fully in \S8.

We summarize in Figure 15 the effects of galaxy separation, subdwarf correction
and Malmquist bias corrections upon the 
density distribution in SA101, one of our deepest program fields.  The affects of bias correction
upon the density profile are subtle and primarily at the extreme ends of the data.

\subsection{Binarism}

It is well established that a substantial fraction of the Population I stars in the Milky Way are not single
stars but are in binary pairs, with estimates for the binary fraction ranging from
50\% in G-type stars (see, e.g., Duquennoy \& Mayor 1991) to 30-35\% in K- and M-dwarfs (Fischer \& Marcy 1992; 
Reid \& Gizis 1997).  Even a binary with an extraordinarily large separation (500 AU) will be 
enclosed within a 1" seeing disk at distances greater than 500 pc and thus will be imaged as a single star.
The effect of unresolved binaries upon our analysis is not intuitive.  Binaries manifest
two observational effects:  The increased apparent luminosity of the ``star" and 
the shift in apparent color.  The former will cause the distance to the ``star" to be underestimated while
the the effect of the latter will vary depending on the relative brightness
and temperature of the two stars.

To examine the effects of binarism upon our survey, we have simulated a density analysis that
includes the effect of binaries.  We adopted the $M_R=f(R-I)$ relation of equations 2 and 3 and
generated a sample of ``stars" with colors chosen from that relation perturbed by Gaussian dispersion 
in $R-I$ and $M_R$ and a density distribution matching a set density law 
(a 300 parsec exponential for simplicity).  For a fraction of the 
stars, we added a companion chosen either at random from the luminosity function
or specified to be nearly equal in
mass to the primary.  The color and magnitude of each primary star was then reset to the 
combined characteristics of the binary.  We then evaluated the density distribution of the stars 
using the original $M_R=f(R-I)$ calibration and the inferred density law was compared with the 
input value.

Figure 16 shows the effect of binaries upon the inferred density law for the
following binary star fractions:  
(a) 50\% fraction of binaries, with equal numbers of equal-mass and random binaries,
(b) 50\% fraction of binaries, all random, 
(c) 25\% fraction of binaries, all equal-mass, 
(d) 25\% fraction of binaries, all random.
The net effect of ignoring binaries is to steepen the derived density law compared
to the true density law.  If we assume a binary fraction of 50\%, then the {\it inferred} scale height in a 
photometric parallax evaluation is approximately 80\% of the {\it actual} value.  The small difference between
simulations with equal-mass and random binaries show that the actual composition of the binaries 
has only a small impact on this evaluation.

One should use caution, however, in applying this correction, as
there is some controversy on the binary
fraction of halo and thick disk stars and whether this binary fraction changes with
spectral type (see discussion in Majewski 1992, \S5.3; Fischer \& Marcy 1992; 
Reid \& Gizis 1997).  For simplicity, our analysis proceeded without this correction.  However, our final
results list both uncorrected values, which are effectively lower limits, and values corrected for a
50\% binary fraction.

\section{Analysis Methods}

\subsection{The Assumed Density Law Forms}

We have used a family of standard density laws to describe the populations of the Milky Way.  A more 
thorough analysis of this subject can be found in Paper I.  We summarize that discussion here.

Disk structures are usually parameterized in cylindrical coordinates by separable radial and vertical
exponentials,

\begin{equation}
\rho(z,r) = \rho_{0} e^{\frac{-z}{Z_0}} e^{\frac{-r}{R_0}}
\end{equation}

\noindent where $z$ is the distance from the midplane, $r$ is the planar distance from the Galactic Center, 
and $Z_0$ and $R_0$ are the scale height and length respectively.  The coefficient $\rho_0$ is the normalizing factor, 
calibrated to produce the observed local stellar density of the disk population.

Exponentials have the benefit of being easily fit in distance-log(density) space where they form straight 
lines.  The slope and y-intercept yield scale height and normalization respectively.  A similar form uses 
the $sech^2$ function to parameterize the vertical distribution.

\begin{equation}
\rho(z,r) = \rho_{0} sech^2(\frac{-z}{2Z_0}) e^{\frac{-r}{R_0}}
\end{equation}

\noindent This functional form has three advantages.  First, it avoids a singularity at $z=0$.  Second, its has a 
more firm theoretical basis in that stars in an isothermal sheet should have such a distribution 
(Camm 1950, 1952; van der Kruit \& Searle 1981).  Finally, at large distances, the $sech^2$ function 
approximates the observed exponential density profile.

Starcounts have retained the exponential formalism almost exclusively despite the fact 
that surface brightness studies of edge-on galaxies have been using $sech^2$ formalism 
for nearly two decades.  While a switch to 
$sech^2$ seems physically justified, there is growing evidence that the 
midplane luminosity of edge-on galaxies {\it is} sharply peaked (de Grijs et al. 1997) and 
exponential functions may indeed provide more reasonable midplane fits than once thought.  In addition, 
Hammersley et al. (1999) have shown that an exponential distribution, despite the 
singularity problems, is a much better fit to the infrared starcounts 
of our own galaxy (although Gould et al. 1996 show the opposite result in optical starcounts).
Given the reasonable observational evidence on both sides, we have 
chosen to try both vertical distribution functions although the differences should be 
relatively minor in our program.

Spheroid population density laws come in numerous formulations.  The most common is the de 
Vaucouleurs (1948) spheroid used to describe the surface brightness profile of 
elliptical galaxies.  This law has been deprojected into three dimensions via the 
Young (1976) formulation as:

\begin{equation}
\rho(R_g) = \rho_{0} exp[-7.669(R_g/R_e)^{1/4}]/(R_g/R_e)^{0.875}
\end{equation}

\noindent where $R_e$ is the effective radius or half-light radius and $R_g$ is the Galactocentric 
distance in spherical coordinates.  This formulation is valid as long as $\frac{R_g}{R_e} > 0.2$ which is true
in all our fields 
for $R_e < 30$ kpc.  Other models have used the power law formulation:

\begin{equation}
%$\rho(R) = \rho_{0} \frac{a_{0}^{n} + R_{\sun}^{n}}{a_{0}^{n}+R^{n}}$
\rho(R_g) = \rho_{0} \frac{1}{a_{0}^{n}+R_g^{n}}
\end{equation}

\noindent where $a_{0}$ is 
the core radius (an often omitted parameter).  The power law formalism, 
beside being a convenient fitting form, is similar to the distribution predicted by cold dark matter (CDM) 
simulations (see, e.g., Navarro et al. 1997).  The CDM formalism is, however, slightly
more complex.  An adaptation to this context is:

\begin{center}
$\rho(R_g) = \frac{\rho_{0}}{(R_g/R_s)(1+R_g/R_s)^2}$
\end{center}

\noindent where $R_s$ is the scale radius.  This differs from the straight power law formalism
in that the slope of the density law will be shallower in the inner regions of the spheroid
than the outer regions.  It should be noted, however, that the stellar halo and dark matter halo are not 
necessarily the same structure.  The dark halo must have a shallower density law ($R^{-2}$) and larger 
mass than the stellar halo to account for the kinematics of the halo.  Thus, formalisms based on CDM 
simulations are of questionable value for starcounts.

In spheroid formalisms, $R_g$ is not true galactocentric radius.  It is corrected 
for the axial ratio $\frac{c}{a}$ as $R_g^2=\sqrt{r^2+(\frac{z}{(c/a)})^2}$.  

For our analysis, the formal distribution of the bulge spheroid is unimportant.  We lack the low-latitude 
first and fourth quadrant fields required to apply constraints to the bulge spheroid and have therefore adopted a distribution
from the literature of a simple spherical $R_g^{-3}$ power law normalized to
.02\% of the solar neighborhood density.  We have left those parameters fixed in all fits.

The more important spheroidal component in our analysis is the halo.  The choice of halo density law
is somewhat arbitrary since the difference between the de Vaucouleurs law and power law are subtle
when seen through such a roughly ground lens as starcounts (although N97 claim that the de Vaucouleurs
distribution underpredicts the faint counts to a noticeable degree).  We have elected to explore
models that use both prescriptions.

The thick disk and halo density normalizations are given in comparison to the density of 
the thin disk at the Sun.  This definition of normalization is slightly confused in the literature with 
some sources normalized to the percentage of {\it total} stars in the Solar neighborhood that belong to that 
population, while others normalize to the {\it ratio} of thick disk or halo stars to thin disk stars.  In this 
contribution, we shall use the latter definition although the difference is only important at high
normalizations.

The normalization to the solar neighborhood naturally must account for the Sun's location
away from the Galactic center and slightly above the Galactic midplane.  We have taken the values for solar radius and 
height to be 8.0 kpc (Reid 1993) and 15 pc (Yamagata \& Yoshii, 1992; N97), respectively.  The
latter value is consistent with the determination of Humphreys \& Larsen (1995) if the thin disk has a 
low (250-300 pc) scale height and this value also corresponds to minima in our $\chi^2$ fitting of the population models.

\subsection{Searching Parameter Space}

The use of photometric parallaxes allows us to make a direct evaluation of the spatial density law.  Rather than
try to fit the structure of the Galaxy in the observed parameter space of colors and magnitudes, we translate the 
observations to discrete density 
measurements at various points in the Galaxy.  This allows us to reduce measurement of the density distribution to the comparatively simple
process of fitting the distance-density distribution simultaneously along our seven lines of site.  In
so doing, we
are essentially fitting the density distribution over the two-dimensional Galactic $r-z$ surface.  In our
case, we assume azimuthal symmetry although we could generalize the formalism to account for triaxiality.  The 
goodness of fit parameter is straightforward - a $\chi^2$ value of the predicted trend
in density-distance space against the actual data.

Finding the $\chi^2$ minimum proved problematical.  The standard grid-search
and gradient algorithms became trapped in local minima.  After many attempts with more elegant 
methods, we elected to take a straight-forward
approach to finding the best density law: Generating millions of density distribution 
combinations with variations of the population parameters over a wide grid of possible values.  Each 
model was 
compared to all seven fields simultaneously and a total seven-field $\chi^2$ was calculated.  The resultant $\chi^2$ values 
were then examined to find minima.  A first pass was made with a broad range of values at coarse 
resolution.  Broad minima were found and further investigations used increasingly finer 
resolution and decreasing range.  To render the problem computationally feasible, subsequent passes varied 
several key parameters while others were held fixed.  Parameters were rotated in and out of variation 
until convergence to a minimum.

\subsection{Binning the Sample}

A concern in our approach is that the von Seeliger equation is a convolution of the
density distribution and luminosity function, which are unlikely to be
independent of each other.  The luminosity function can vary with location in the Galaxy (and population) 
and the density
distribution can vary depending on spectral type.  For example, younger stars tend to be more confined
than older stars in their vertical distribution from the Galactic midplane.
Thus, the problem is not simply two-dimensional (i.e., distance against
density) but {\it three-dimensional} (i.e., distance against density against luminosity).  

Sophisticated models incorporate this convolution effect in their simulations (see, e.g., 
Robin \& Creze 1986; Paper I).  An alternative strategy is to break the data into groups of similar stars.
Ideally, one might evaluate stars of identical {\it mass}.  However, because deriving masses for stars is even more
subject to systematic errors and requires additional degrees of extrapolation, we have elected to 
evaluate stars of identical estimated absolute magnitudes.  Our analysis is 
limited to the
range of magnitudes $5.8 \leq M_R \leq 10.2$.  For solar metallicity stars, this corresponds to 
approximately $6.3 \leq M_V \leq 12$ or spectral types K0 to M4.

\subsection{Convergence and Parameter Drift}

We found that our exploration of parameter space defined likelihood minima quite easily for the 
two disk populations.  The parameters of the halo, however, diverged quickly to edge of parameter
space.  With the power law 
halo, this was manifested as a derivation of low power law indices, low axial ratios and high normalizations 
($\sim$ 1.0, 0.3 and 0.6\% respectively).  The de Vaucouleurs halo also 
diverged to low axial ratios, high effective radii and high normalizations ($\sim$ 0.3, 10 kpc and 0.5\% 
respectively).  With the halo parameter drift came a commensurate secondary effect of flatter thin and thick disks 
of low scale heights (150 and 500 pc respectively).

It is, of course, possible that those results are the correct description of the Galaxy.  Our
analysis was designed specifically to ignore {\it a priori} assumptions about the populations and 
find the model that best fits the data, regardless of how it compared with previous results.  However, 
the derived parameters are so at variance with the prior literature that one is led to believe that 
a fundamental flaw is present in the method.

The likely explanation for this divergence is the general insensitivity of starcounts to halo 
parameters, a problem magnified by the crudity of our analysis of photometric parallaxes.  It is also possible that the 
halo has a more complex field to field variation than the simple models parameterized above.  Our survey 
would be among the first starcount surveys to be complete enough and deep 
enough to encounter problems of halo substructure affecting an analysis of starcounts.  An 
underconstrained complex halo density distribution could easily be divergent in such an analysis.  
The case for a complex halo is made in detail in \S8.

\section{Best Fit Conventional Models}

\subsection{Disk Models With Constrained Halos}

Given the failure of any of our models with an unconstrained halo to converge, we decided to adopt fixed halo 
distributions from the existing literature.  We then allowed the thick and thin disk parameters to 
vary freely (except for the thin disk scale height in the intrinsically brightest stars, which are 
undersampled close to the sun) to see which provided the best description of the Galaxy.

Models for the halo population were taken from L96, N97, B99 (which is similar to the Paper I halo) and R00 
(the flattened version of which is identical to the halo derived in C01).
The results are listed in Table 3.  The spherical halos adopted by N97 and B99 produce 
dramatically poorer fits than flattened halo models.  This is illustrated in 
Figure 17, which compares the interim model of Paper I against the isodensity points for our 
brightest ($5.8 \leq M_R < 6.8$) stars along our seven lines of sight.  These stars provide the greatest
leverage on the halo because they probe the farthest into the Galaxy.  We can immediately discern that the 
interim model is a good description of the density distribution at the SGP and a reasonable, but
imperfect description of the density law outside of $R_g$=5 kpc.  On the other hand, 
there are significant density overpredictions toward the Galactic Center in the distant bins
of SA107/SA184.  These overpredictions are also seen in the less-flattened Besan\c{c}on model to a lesser 
but still noticeable degree.

The best halo prescription appears to be either the L96 flattened de Vaucouleurs spheroid or the 
Besan\c{c}on/C01 flattened power law.  In fact, the predictions of the two models are similar.  
Both produce a low scale height, high normalization thick disk.  Table 4 lists the best fit to all 
four absolute magnitude data bins given the Besan\c{c}on halo formulation and exponential disks and Table 5 
gives the best fit using 
$sech^2$ disks.  Note that the values in Tables 
3-5 are not corrected for binarism.
The errors bars are taken at the point at which $\chi^2$ = min 
($\chi^2$)+1.  The asymmetry in the disk scale length error bars is caused by the ease with which sharp
density gradients can be ruled out in comparison to shallow gradients.  The $\chi^2$ interval of 2.5-3.0 
encapsulates any power law halo with an axial ratio 
of 0.5-0.7 and a power law index of 2.5-3.5, with low index, low axial ratio halos favored.  The 
thick disk that results from any of those halo prescriptions has the general parameters of 
$Z_o$ = 700-900 pc, $R_0$ = 3000-5000 pc, $\rho$ = 6-10\%.  It is curious that the lowest absolute
magnitude bin has a significantly higher thin disk scale height.  This could reflect either
that scaleheight increases at fainter magnitudes or, possibly, that binary fraction declines at 
fainter magnitudes.

Figures 18 and 19 show the overplot of the observed isodensity points on the model isodensity 
contours for our ``best-fit" models.  While those two models are a reasonable description of the 
Galaxy, especially for nearby stars, there remain systematic discrepancies for the most distant 
density contours.
In particular, the models overpredict the density toward the Galactic Center and 
underpredict the density in the outer halo regions.  The systematic, as opposed to random, 
nature of the over- and underpredictions, indicates that the standard models are still in need of 
revision.

\subsection{The Shape of the Galaxy}

Table 6 lists the revised parameters of our new model.  The parameters are a general 
synthesis of the magnitude-differentiated descriptions listed in Tables 3-5 and reflect the range of 
values occupying the $\chi^2=2.5-3.0$ region of parameter space.  The listed parameters primarily pertain
to the bluer stars and do not reflect the increased thin disk scale height at faint absolute magnitudes.
Both raw values (i.e, uncorrected for binarism) and corrected values (with a binary fraction of 50\%)
are listed.

An old disk with an exponential scale height lower than the canonical value (325 pc) has been 
indicated in much of the recent literature, with the exception of C01.
We have only parameterized the thin disk with a single old population, which may be too simplistic 
(for example, N97 uses three fixed disks).  Our data near the Galactic 
midplane are too sparse to constrain 
more than one thin disk component.  In 
addition, at the faint absolute magnitudes that our study explores, the thin disk is dominated by the old disk 
component (see Paper I).  The equivalent 
$sech^2$ scale height places the Galaxy at the somewhat thin end of the spectrum of spatial 
parameters of edge-on galaxies (de Grijs 1998).

Our thin disk scale length of 2.25 kpc is shorter than the 3-4 kpc canonical value, but the error 
bars enclose the longer estimates of R96, L96 and B99.  A short thin disk scale length is also indicated by 
infrared starcounts (see, e.g., Ruphy et al. 1996; Drimmel \& Spergel 2001).
Studies of edge-on disk galaxies have found the average $\frac{R_0}{Z_0}$, in which $Z_0$ is defined 
slightly differently as the scale height in the density formulation 
$\rho(z)=sech^2\frac{z}{Z_0}$, to be approximately $5.9 \pm 0.4$ (de Grijs \& van der Kruit 1996).  
The axial ratio is dependent on Hubble type (de Grijs, 1998) and ``super-thin" galaxies 
with axial ratios as high as 14 have been found (Matthews 2000).  The Galaxy's axial ratio of 
$\sim 4$ places it among the Sc galaxies.

Our derived thick disk scale height is similar to the C01 result and comparable to the Besan\c{c}on results (R96), albeit
with a noticeably higher normalization than the latter.  Comparing normalizations between different studies is difficult
because some authors are unclear exactly {\it to what} their thick disk and halo are being normalized.  It is possible 
that since both our study and C01 normalize to the old disk alone, we overestimate the normalization 
by ignoring contributions from younger populations that have a smaller signature in our
study.  Based on the three-disk normalizations used in N97, a better normalization 
for our thick disk might be 4-6\%.

The high thick disk scale heights of Paper I, GR83 and other studies are likely the result of
surveying a single direction of sky near a Galactic pole and adopting a fixed near-spherical halo. 
It was concern about halo flattening and degeneracy that motivated
this study -- to determine if non-polar fields demanded a more complex density distribution.  Such concerns
appear to have been justified.
While a 1.5 kpc height, 2\% normalization thick disk is a 
suitable fit to SA141, as attested by Figure 17, the non-polar fields argue strongly
for a much more substantial thick disk population.

Presumably, the spatial distribution of the field stars is a reflection of their dynamics.  Any self-consistent
dynamical Galaxy model should simultaneously account for both.  The measured $\sigma_W$ 
of the thick disk is approximately 40 km s$^{-1}$ 
(Norris 1986; Sandage 1987; Sandage \& Fouts 1987; Carney et al. 1989; Beers \& Sommer-Larsen, 1995; 
Guo 1995; Reid et al. 1995; 
Ojha et al. 1996; Chiba \& Beers 2000), which implies a vertical scale height 
of around 1 kpc, with possibly as much as 10\% of the local stars included
in this high $\sigma_W$ population (Sandage \& Fouts 1987; Sandage 1987; Casertano et al. 1990; 
Reid et al. 1995) although Guo (1995) argues for a lower normalization of 3\%.  While the thick disk 
scale height implied by these studies is roughly compatible with our results, it is also 
roughly compatible with just about every measure of thick disk scale height that has been made 
(see Table 1).  A fraction of local stars in the high $\sigma_W$ population as high 
as 10\% would be compatible with our own results and 
those of R96 and C01.  Finally, Majewski (1992) and Guo (1995) have argued that
thick disk kinematics dominate out to at least $Z=4.5-5.5$ kpc.  In our model, 
the thick disk dominates the density distribution to a similar height above the plane.

Our thick disk scale length is compatible with previous starcounts results as well as the kinematical
estimate of Chiba \& Beers (2000).  More important than the precise value of the thick disk scale length, 
however, is that the thick disk scale length is longer than the thin disk scale length, at least when evaluated 
in our brightest absolute magnitude bin -- which has the smallest
uncertainty.  Although it has been standard practice in the literature to assume that the two lengths should 
be equal or comparable, in many starcounts studies, including this one, the error bars on 
the scale lengths are too large for firm conclusions to be drawn.  L96 is a notable exception, showing a 
thick disk scale length longer than the thin disk at a $5 \sigma$ level of significance.  In addition,
studies of edge-on disk galaxies have shown that some of them (especially early-type galaxies) exhibit 
radial variations in scale height, indicative of thick disks with scale lengths longer than the thin disks 
(see, e.g., de Grijs \& Peletier 1997).

\section{Toward More Complex Density Distributions}

There could be a number of reasons why we are unable to fit a standard density law to the halo starcounts 
and why even the best models from the literature have systematic problems.  These divide into two 
categories:  Problems with our analysis and problems with the density law itself.

Our analysis could be compromised by a poor subdwarf correction.  This 
would produce systematic effects.  While our simplified metallicity correction fit the data on 
subdwarf parallaxes reasonably well, the number of stars with good parallaxes and good photometry is 
small and the scatter large.  There is a critical need for high precision parallaxes and broadband
photometry of metal-poor 
stars, a situation that could be remedied by ground-based photometry and the astrometric 
space missions of the next decade.  Figure 14, however, shows that while the subdwarf correction is 
large for the latest stars, it is small for the early K dwarfs that provide our only probe into the 
halo.  Unless the subdwarf correction is grossly incorrect, this would not produce the effects we 
are seeing.

Moreover, the apparently small subdwarf correction in blue stars may itself asymptote at low abundances so that 
even significantly poorer halo populations would have similar absolute magnitudes to [Fe/H]=-1.5 
stars.  This ``guillotine effect", in which the metallicity correction asymptotes as metallicity 
declines, has been known at least since first defined in the context of ultraviolet excesses by 
Sandage (1969).  We do not have nearly enough subdwarfs with parallaxes to characterize this effect 
and the scatter of the red [Fe/H]=-2.0 stars in Figure 14 cautions against attempting to define such
an effect.  Nevertheless, the possibility of a 
guillotine effect -- especially in the bright blue stars upon which our analysis
leans heavily -- mitigates to some extent the danger of subdwarf correction errors for
very low metallicity populations in the Galactic halo.

Related to the subdwarf problem is the possibility of an incorrect metallicity gradient.  The Yoss 
et al. metallicity gradient is poorly constrained in the halo.  While it is generally agreed that 
[Fe/H]$\sim$-1.5 is a characteristic halo metallicity, the halo has a high metallicity dispersion and may, 
in fact, have some metallicity substructure (Carney et al. 1996; King 1997).
In addition, none of our models have incorporated a {\it radial} metallicity 
gradient but only a vertical variation.  This could be too simplistic, although a 
radial gradient of increasing metallicity toward the Galactic Center would cause systematic biases 
producing the opposite effect of the observed density overpredictions.  Moreover, 
radial metallicity gradients have only been demonstrated for the thin disk (Rong et al. 2001), a population that is
minimally represented in our most distant stars.
Metallicity scatter or substructure would presumably not induce first order systematic effects.

A final source of error could be an overzealous galaxy separation effort.  The broadening of the 
stellar $\chi$ and SHARP loci at faint magnitudes could cause faint stars to be preferentially 
misclassified.  Such a problem, however, would result in a {\it uniform overprediction} of 
density at faint magnitudes, not the selective over- and underprediction that is observed to occur.

The second possibility is that the density formulations we have used, while reasonable
descriptions for the local Galaxy, are too simple to describe the Galactic density distributions 
at great distances from the Sun.  There are several density law modifications that could resolve the 
observed problems:

1.  {\it Thick Disk Flaring.}  Our models have assumed that the thick disk scale height
is constant with radius from the Galactic center.  This formulation has only been tested
in the Galactic context to rather short distances and primarily toward the 
Galactic anti-center.  It is possible that the thick disk is flared, with a scale height that
{\it increases} with radius.  This would help eliminate some of the density overprediction in SA107/SA184.

The remaining five fields in our survey cover a small dynamic range in planar distance.
When we evaluate these five fields independently of SA107/SA184, the resultant uncorrected thick disk
scale height is $\sim$ 0.8-1.2 kpc, although this value is highly dependent upon the adopted
halo model.  This suggests that the local thick disk scale height may indeed be larger than
the scale height in the inner Galaxy.  Alternatively, this result could indicate that the
radial density distribution of the thick disk is not exponential, as we have assumed.  A thick disk 
that is exponential locally but that has a shallower radial density gradient in the inner Galaxy would 
also fix the inner Galaxy isodensity contours.

2. {\it Triaxiality.}  Larsen \& Humphreys (1996) have argued for a triaxial halo based upon 
asymmetries found in halo starcounts from APS-POSS.  Our data cannot detect this proposed 
distribution because we do not have fields in the lower-latitude areas noted to have asymmetry by 
Larsen \& Humphreys.  This would also leave unexplained the overpredictions of the model in the SA107/SA184 
fields, which would show minimal effects from the Larsen \& Humphreys asymmetry given the location of these
two fields near the 
Galactic meridian and would show opposite effects (i.e., overprediction in one, underprediction
in the other) if they were affected by a halo asymmetry.

3. {\it A Complex Halo.}  
Evidence has been growing for some time that simple descriptions of the Galactic halo are inadequate.
Studies of the outer halo through either giant stars or bright halo stars have derived a 
spherical $R^{-3}$ distribution (Paper I; N97; Majewski et al. 2000; Morrison et al. 2000; 
Majewski et al. 2002).  Studies of fairly nearby main sequence stars (Larsen \& Humphreys 1994; R96; 
L96; R00; C01) and microlenses (see, e.g., Samurovic et al. 1999) have derived a flattened halo 
distribution.  Some surveys have found a single axial ratio too restrictive and have adopted a $\frac{c}{a}$
ratio that increases with Galactocentric radius to explain the spatial distribution of
RR Lyrae/HB stars (Hartwick 1987; Preston et al. 1991; Kinman et al. 1994; 
Layden 1993; Layden 1995; Wetterer \& McGraw 1996; Sluis \& Arnold 1998), globular clusters 
(Zinn 1993) and main sequence stars (Gilmore et al. 1985).

Studies of the kinematics and abundance of 
both field stars and globular clusters show that the halo is better 
described as having {\it two} subpopulations -- a flattened 
inner subpopulation with a metallicity gradient and slow-rotation kinematics and a round outer 
subpopulation with no metallicity gradient and anisotropic kinematics (Zinn 1993; Dinescu et al. 1999).
Recent studies of halo field stars selected either by proper motion (Carney et al. 1996) or 
metallicity (Sommer-Larsen \& Zhen 1990; Allen et al. 1991; 
Chiba \& Beers 2000) have presented a similar picture -- a flattened inner halo -- and a round 
$R^{-3}$ component that lacks a metallicity gradient.  Additional support for dual halo models can be drawn 
from the apparent dichotomy in detailed chemical abundances of halo stars (Nissen \& Schuster 1997).

Starcounts studies have never explored this possibility.  Such models might resolve many of the 
disagreements in starcounts results.  In a dual halo model, nearby stars (R96; L96; C01) are dominated 
by the flattened inner halo while distant stars (Paper I; N97) are dominated by the the spherical 
outer halo.

Adopting a two halo formalism, however, leads to a poor solution.  If the 
$R^{-3.5}$ spherical power law for the outer halo is extrapolated into SA107/SA184, the result is a 
severe overprediction in the counts.  This can be mitigated by expanding the core radius of the the 
outer halo power law or lowering the power law index.  That change, however, results in a density 
gradient in our high-latitude fields that is far shallower than the observed gradient.  We have thus 
far been unable to find a uniform model that correctly predicts all of the fields, although there are 
several avenues that show promise for future work.

In particular, in an effort to fit all of the density
contours, we attempted three variations of the standard halo that showed some improvement over conventional 
models.  Our first used a power law
in which the power law index increases with radius.  This was motivated by the results of
recent hierarchical clustering models of galaxy formation, which indicate
that the power law indices of accreted dark matter halos should fall toward the inner regions, 
producing a shallow $R^{-1.5}$ inner halo density gradient and a steep outer $R^{-3}$ to $R^{-4}$ outer 
halo density gradient (see, e.g., Dubinski \& Carlberg 1991; Navarro et al. 1997; Subramian et al. 
2000; Dav{\'e} et al. 2001).  While the metal-poor stellar halo discussed here is a completely 
separate entity from the (presumably) primordial dark-matter halo, the density law of the former should, 
after all, follow the gravitational potential of the latter.  If that were the case, 
distant field stars of the stellar halo
could offer a means of testing dark-matter models.  Our second variation used 
confocal ellipsoids to describe the halo.  This
projection produces elliptical inner isopleths and round outer isopleths.  Our final model attempted to
use the $\frac{c}{a} = f (R)$ formulation of Preston et al. (1991) to account for axial ratio changes.  
All three halo formulations produce measurable improvements to the {\it initial} fits but 
do not converge, which may once again reflect the general insensitivity of starcounts to halo 
parameters -- an insensitivity that is only worsened when more free parameters are added.

On the other hand, it is possible that while a dual halo description of the Galaxy is qualitatively
accurate, the very nature of the model renders it unamenable to 
conventional quantitative analysis.  The spherical
outer halo {\it may not have a smooth density distribution} but be comprised of overlapping streams 
of stars.  Substructure and/or breaks in the halo density profile have been directly observed in 
giant stars (Majewski et al. 2002), main-sequence stars (Majewski et al. 1994, 1996; Newberg et al. 2001; 
Dinescu et al. 2002)
and BHB/RR Lyrae halo stars (Yanny et al. 2000; Ivezic et al. 2000; Vivas et al. 2001).  These studies have suggested
that at large Galactocentric radii, the halo may be largely or {\it entirely} comprised of overlapping 
streams and that smooth density distribution may only apply to the inner Galaxy.

It is unclear if a substructured outer halo could resolve the remaining problems in our model.
The deviations from our model appear to be 
{\it systematic}, rather than the {\it random} deviations one would see if looking out through a 
``can of worms".  Additionally, Johnston (1998) 
has indicated that even major accretion events might be undetectable in starcounts studies. 
If the filling factor of the star streams is high enough, what appears as  a ``can of worms" 
in the finely tuned studies of bright tracer stars could be blurred into a vague, indistinct,
seemingly uniform population in starcount studies.  Moreover, because 
we have only seven fields in our program, our ability to distinguish between systematic and random
effects in the outer halo is limited.

If a flared thick disk were incorporated to reconcile the overpredictions of SA107/SA184, however,
halo substructure could explain the underpredictions in our other fields, particularly SA141, which
appears to be in the direction of an outer halo stream of Newberg et al. (although the stream appears to be at 
larger distances than we are sensitive to).  Thus, the idea of a dual halo in
which the outer halo is not a uniform structure but a vague ensemble of star streams, remains an 
intriguing possibility and a likely resolution of the outstanding discrepancies in our model.

\section{Implications for Galactic Formation}

Starcounts are, of course, a means to an end.  The actual numerical values of 
the various parameters of the populations are less scientifically important than what they 
tell us about the Galaxy in general -- for example, the origin of the various components.

A number of scenarios have been proposed for the origin of the thick disk (see review in Majewski 
1993).  It is clear from its main sequence turnoff color that the thick disk includes few, if any, main-sequence stars
younger than a few Gyrs.  At present, a popular formation mechanism is that the thick disk formed by tidal heating
of the early Galactic thin disk either by merging galaxies (see, e.g. Quinn et al. 1993, 
hereafter Q93; Walker et al. 1996; Huang \& Carlberg 1997; Sellwood et al. 1998) or the late infall of 
stellar clumps formed during the dissipational collapse of the Galaxy (Noguchi 1998).  In tidal 
heating or merger scenarios, the Galactic gas content would be kinematically heated by the event but 
would eventually cool off and settle down to form the present thin disk.  Because stars are 
effectively collisionless particles, however, they would retain the dynamical heat of the event.  

Alternatively, the thick disk may have formed in a dissipative collapse following
the rapid formation of the halo and ending in the formation of the thin 
disk (see, e.g. Larson 1976; Jones \& Wyse 1983).  Majewski (1993) outlined evidence to support a
variant scenario in which the thick disk represented the initial structure formed in the Galaxy via a 
global collapse as first theorized by Eggen, Lynden-Bell \& Sandage (1962, hereafter ELS), and this was 
followed by the formation of the halo by accretion of small stellar systems (see also Sandage 1990).

Evidence in favor of a heating origin for the thick disk comes from studies of 
chemical abundances (Nissen \& Schuster 1991; Fuhrmann 1998; Gratton et al. 2000; Prochaska et al. 
2000) and kinematics (Carney et al. 1989; Beers \&
Sommer-Larsen 1995; Ojha et al. 1996), which seem to show a continuum of properties 
from nearby halo to thick disk and then a discontinuity between the thick
disk and thin disk.  One
must always temper claims for such discontinuities with the possibility of selection bias.  Thick
disk stars are often selected for kinematical study by their chemical properties or for chemical study by 
their kinematics.  In a late-heating scenario, 
the most extreme kinematical and most extreme chemical portions of the thick disk could be the
same.  The more complete samples of globular clusters (Zinn 1993) and field 
stars (Majewski 1992, 1995; Chen 1997; Chen 1999; Chiba \& Beers 2000) show kinematical gradients in the 
disks and a kinematical break between thick disk and halo.  In addition, the time of such a violent heating event can be constrained to be 
between the age of the thick and thin disk stars.  It therefore must have occurred fairly early in 
the Galaxy's history since the oldest stars of the thin disk are between 8 and 12 Gyr in age 
(Janes \& Phelps 1994; Bergeron et al. 1997; Leggett et al. 1998; Jiminez et al. 1998; Wood \& Oswalt 
1998; Knox et al. 1999; Carraro et al. 1999; Montgomery et al. 1999; Liu \& Chaboyer 2000).  The 
thick disk stars are generally thought to be older than this (Gilmore \& Wyse 1987; Carney et al. 
1989; Rose \& Agostinho 1991; Gilmore et al. 1995).  One important caveat, however, is that the differences in
ages (generally around 1-2 Gyr) are small compared to the uncertainties.  Given the difficulty in measuring
the age of field stars, the age difference between the thin and thick disk is still
debatable.

In our opinion, the most compelling recent evidence for a merger origin is that
the star formation history gleaned from abundance patterns does not conform
to expectations from dissipational formation  (Prochaska et al. 2000), although this evidence comes with the 
caveat that the Prochaska et al. thick disk stars do not have a large abundance overlap with the thin disk stars
to which they are being compared.
There is also strong support from the extragalactic angle in that thick disks 
are not seen in all edge-on disk galaxies (van der Kruit \& Searle 1981; Morrison, Boroson \&
Harding 1994; Morrison et al. 1997; Fry et al. 1999), which suggests a more stochastic process is at
work.

The origin of the halo is thought to be either through the rapid global collapse of the protogalactic
gas cloud (ELS) or the accretion of protogalactic fragments (Searle \& Zinn 1978).  The
accretion hypothesis is in line with favored
cold dark matter cosmologies, which predict a hierarchical galaxy formation process
(see, e.g., White \& Rees 1978; Navarro et al. 1997).  \S8 recounts
the evidence on both sides of the halo formation question and notes the attempts to reconcile the observational
evidence by dividing the halo into two distinct populations.

Four Galactic populations (old disk, 
thick disk, inner halo and outer halo) can account for most of the properties of
Galactic field stars found in the literature.  The outer 
halo -- as described by Zinn (1993) and Chiba \& Beers (2000), described as just ``the halo" by 
Majewski (1993) and completely absent from studies of comparatively nearby main sequence stars (e.g., R96, 
L96, R00) -- is more or less spherical, shows phase-space substructure, is chemically inhomogenous and 
disjoint from 
the inner halo.  It is the product of Galactic accretion of globular clusters and dwarf 
galaxies.  The inner halo -- as described by Zinn (1993) and Chiba \& Beers (2000), difficult to 
distinguish from the thick disk in the Galactic poles and therefore blurred with the thick disk 
into ``IPII" by Majewski (1992, 1993, 1995) and Paper I, described as ``the halo" in R96, L96, R00 and 
C01 -- is flattened, has kinematical and metallicity gradients and is chemically and kinematically overlapped
with the thick disk.  It is likely the product of an ELS-like collapse (Sandage 1990).  The thick disk is then 
the compact, chemically and kinematically homogenous structure described in R96 and C01 and 
Prochaska et al. -- possibly formed by a merger
early in the Galaxy's history.

Such hybrid models for the formation of the halo have been
proposed (see, e.g., Sandage 1990; Majewski 1993; Norris 1994; Chiba \& Beers 2000).
The principal contribution of starcounts in constraining Galactic formation scenarios lies in 
revealing the underlying shape, chemistry and ages of the stellar population through
sophisticated modeling, study of detailed stellar color distributions or ultraviolet
excesses -- all of which we plan to pursue with our present data set.
Our study of photometric parallaxes, however, provide some support for the hybrid formation model in that 
we show discrepancies between the relatively simple canonical density laws and the observed density 
distribution of Galactic field stars.  These discrepancies could be resolved by invoking structural 
properties unique to the hybrid model.  For example, the density overprediction in the inner 
Galaxy would be explained by thick disk flaring, which Q93 assert would be characteristic of a 
kinematically heated thick disk.  Additionally, the density underpredictions in the outer Galaxy would be 
resolved by a second spherical substructured halo -- a population characteristic of an 
accreted halo.

The general picture that is emerging in the study of the Galaxy is one in 
which both global collapse and accretion have probably played significant roles in 
producing the extant stellar populations.  Accretion may be
manifested in two modes -- the subsuming of accreted stellar populations 
into the outer halo and the tidal inflation of the early thin disk by a 
particularly large merger.  Global collapse would be manifested in the inner halo.
Continued exploration of our own data set and that of
SDSS will continue to improve our understanding.

\section{Conclusions}

Our photometric parallax survey of seven Kapteyn Selected Areas has produced a number of 
intriguing results:

$\bullet$ The thin disk is well described by a double exponential of raw scale height 280 pc and scale length
2-2.5 kpc.  Its axial ratio is consistent with the ratios observed in edge-on Sc galaxies.  The 
faintest stars in the sample show some evidence of an elevated scale height (350 pc).  Interestingly, 
recent results indicate that the scale height of thin disk white dwarfs may also be much higher than the
canonical thin disk scale height (Majewski \& Siegel 2002; Nelson et al. 2002).  However, if the
binary fraction declines for faint stars, this would produce a similar effect.

$\bullet$ The thick disk is well described by a double exponential of raw scale height 700-1000 pc and scale length
3-4 kpc.  Approximately 6-10\% of the local old stars are part of the thick disk.  There is
some evidence that the disk is flared, which is consistent with a kinematic heating origin
for the thick disk.

$\bullet$ If the binary fraction of Population II stars is 50\%, the scale heights of the thin
and thick disk for blue stars would increase to 350 and 900-1200 pc, respectively.  The elevated thin disk scale 
height of the faintest stars is 440 pc.

$\bullet$ Fits to the halo density law do not converge in our simulations.
Flattened ($\frac{c}{a}\sim0.6$) power law ($\rho \propto R_{gc}^{-2.5}$) halos are generally favored.   

$\bullet$ Some of the remaining discrepancies between our model and the data could be resolved
by a dual halo model in which the inner halo is flattened and the outer
halo is a roughly spherical ``can of worms", consisting of distinct overlapping streams of stars.
Such a dual halo would also reconcile many of the discrepancies in the literature into a single model.

$\bullet$ Evidence for a flared thick disk and a dual halo would support the notion that the Milky Way has grown
at least in part through the accretion of external systems.

Future contributions from this series will explore the new CCD starcount data in greater detail with the goal
of strengthening our
understanding of Galactic stellar populations and the underlying chemodynamical history that they
reflect.

\acknowledgements

This research was supported by NSF grants AST-9412265 award to 
IBT and SRM, grant AST-9412463 awarded to INR, CAREER Award AST-9702521 to SRM, the David and Lucille Packard Foundation 
(SRM, MHS) and the Observatories of the Carnegie Institute of Washington (SRM, INR, IBT).  The authors
thank Robert Link and Jamie Ostheimer for helpful discussions and the numerous Swope
telescope operators for their assistance.

{\bf Figure Captions}

\figcaption[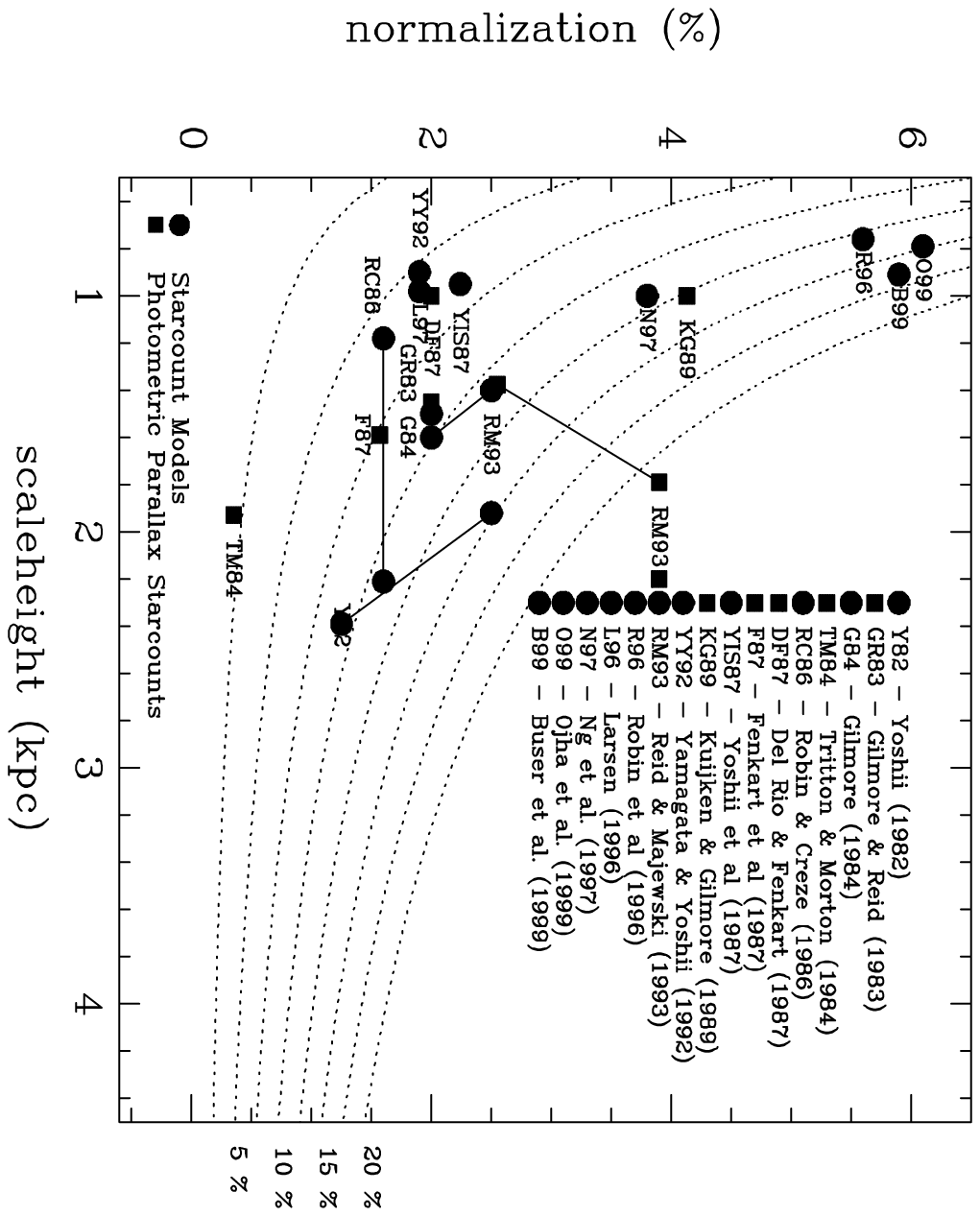]{Previous measures of the thick disk (IPII) exponential density law 
normalizations (abscissa) and scale heights (ordinate).  This plot does not include Chen et al. 2001, 
which is displaced significantly off of the figure.  The dotted lines reflect the percentage of the Galactic stellar mass
residing in the thick disk for various combinations of parameters.  Squares points represent the
results of previous photometric parallax surveys.}

\figcaption[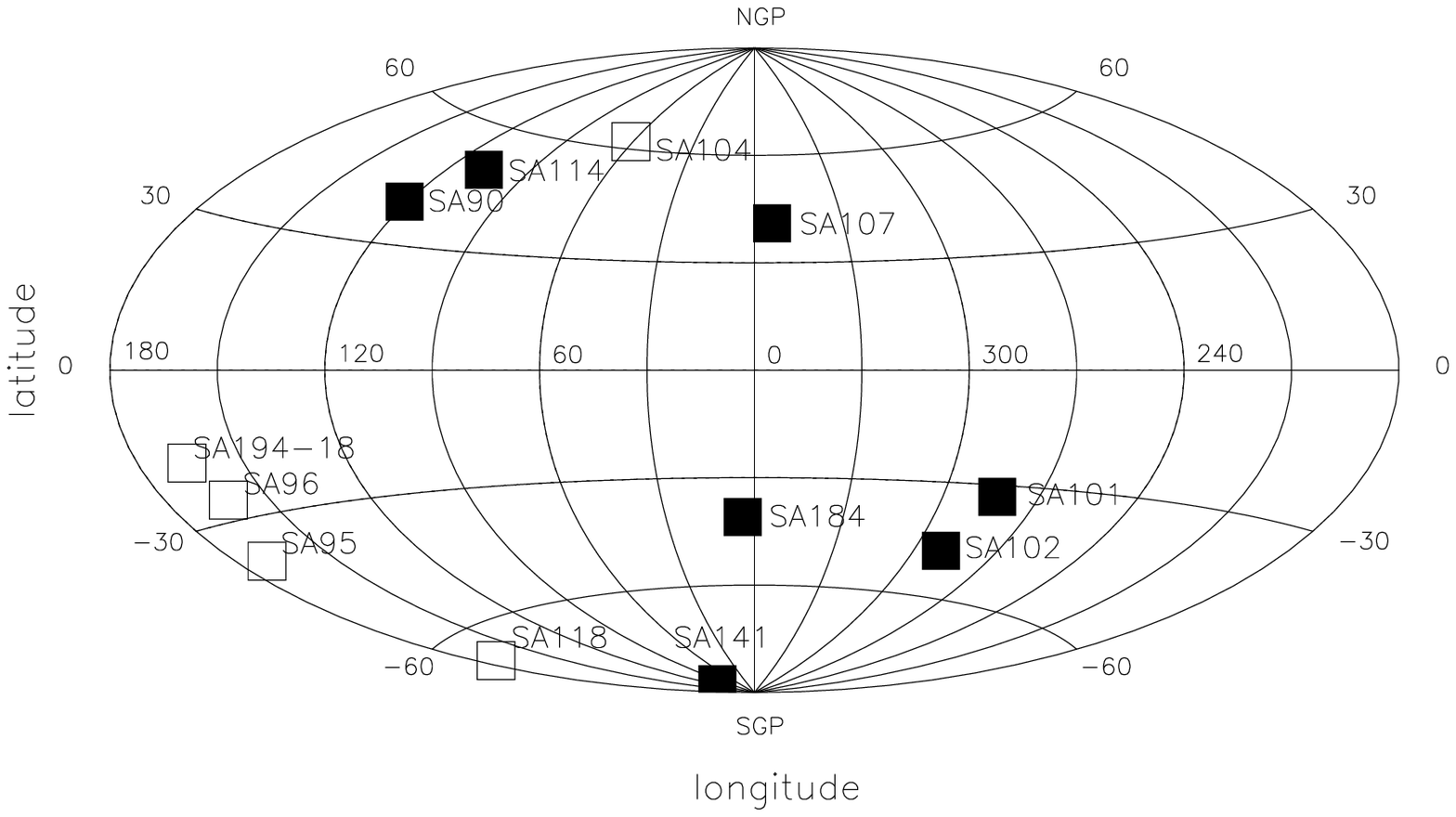]{Our starcount fields in Aitoff projection.  The darkened squares are those
analyzed in this contribution.  Fields represented as open squares have been reduced through the data pipeline
but lacked complete $RI$ data for this particular study.  Later analyses will examine all twelve fields.}

\figcaption[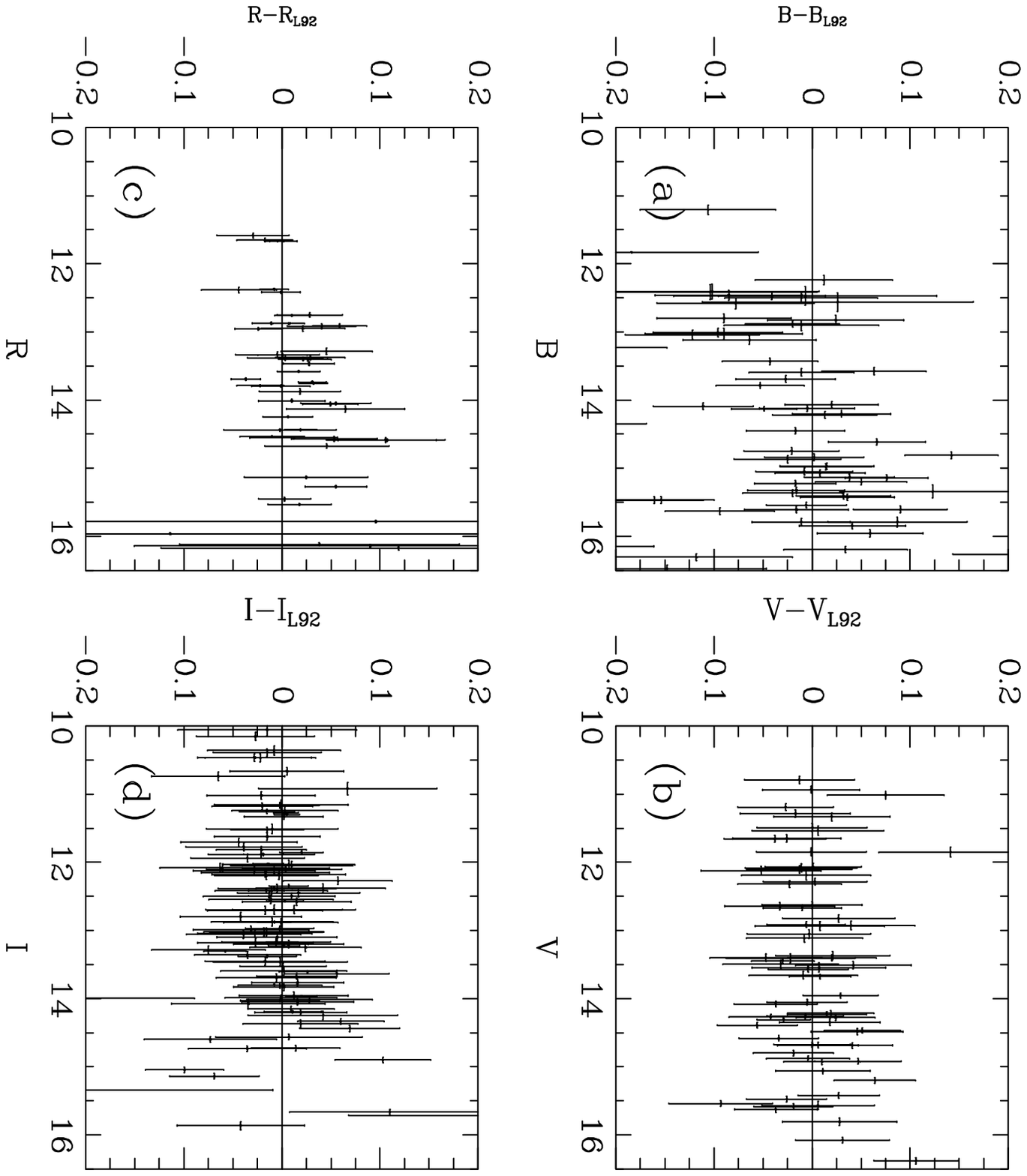]{The magnitude difference of standard stars measured in
our pipeline compared with the photometry of Landolt (1992).  The smaller error bars correspond to the Tek 5 data.
The apparent trend in $B$ is not statistically significant.}

\figcaption[f4.eps]{The magnitude-SHARP and magnitude-$\chi$ distributions of a typical
field, in this example the $V$ band frame of the central field of SA107.  The upper loci in each panel are
non-stellar objects while the main locus near SHARP=0.0, $\chi$=1.0 are stars.  The horizontal lines
indicate the cuts applied to those fields while the vertical lines show the classification limit of the 
field.  More stringent $\chi$ limits were found occasionally to filter out bright stars, such as the 
clump of stars with elevated $\chi$ values around $V=17$ show in the figure.  SHARP is therefore adopted as 
the primary discriminant in our program.}

\figcaption[f5.eps]{The recovered fraction of stars added to a diverse sample of CCD 
images with respect to the classification limit.  Panel (a) shows the overall trend, while panel (b) 
highlights the region near the classification limit.}

\figcaption[f6.eps]{The recovered fraction of stars with respect to the 
classification limit after galaxy decontamination.  Panel (a) shows the overall trend, while panel (b) 
highlights the transition zone.}

\figcaption[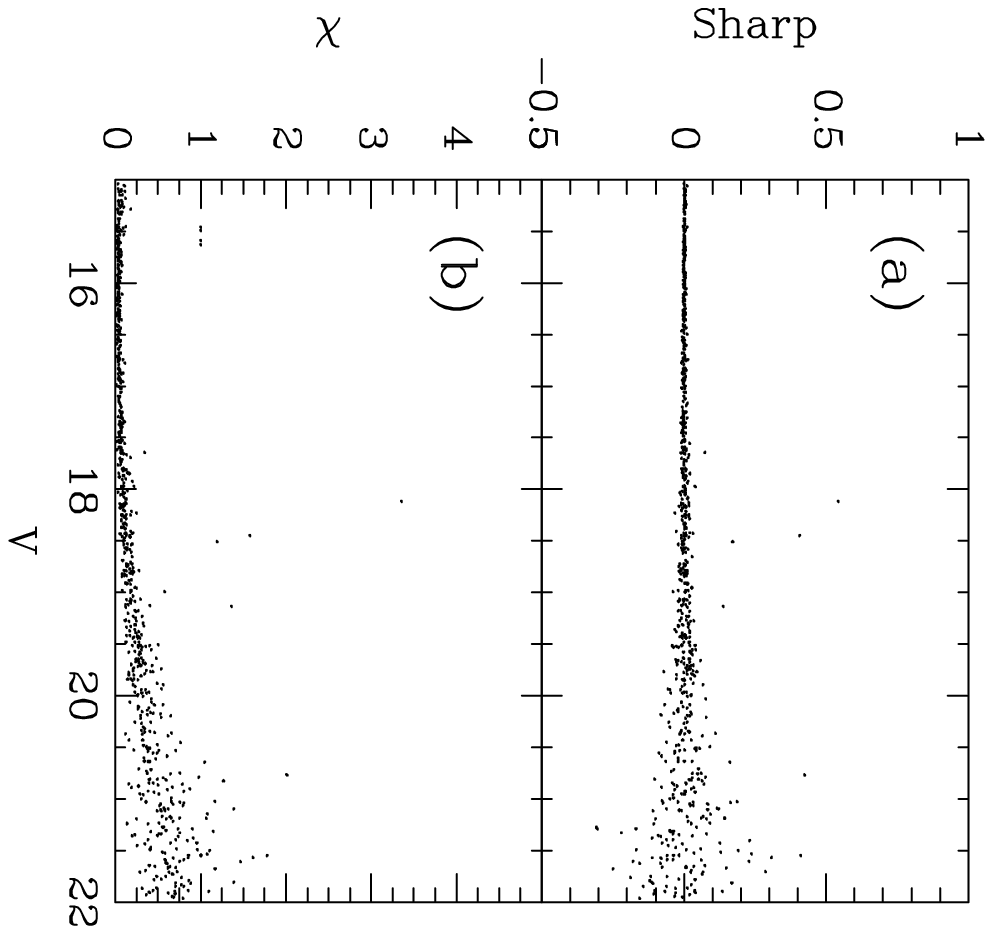]{The magnitude-SHARP and magnitude-$\chi$ distribution of
artificial stars.  This can be contrasted to the distribution of real data seen in Figure 4.  Note the decreased
$\chi$ values for bright stars and the much tighter clumping of the stellar locus.  This is the 
result of unrealistically excellent photometry for artificial stars.}

\figcaption[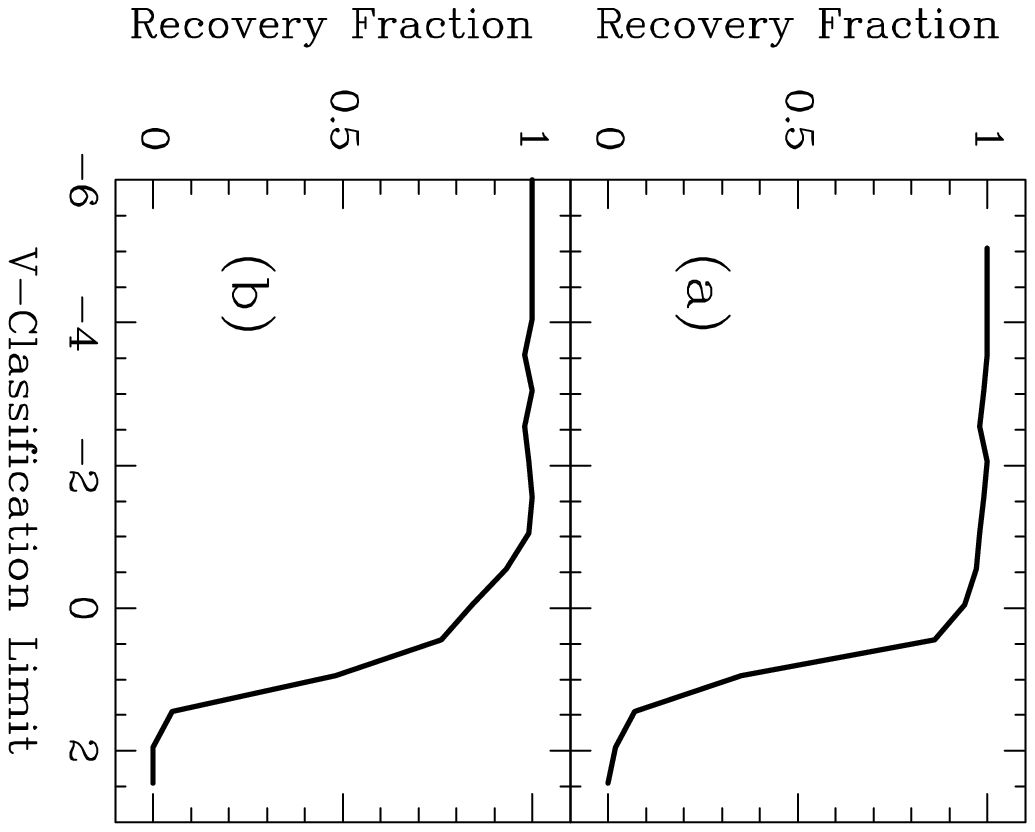]{The recovered fraction with respect to the classification limit of stars 
added to the CCD frames with ARTDATA.  (a) shows the recovered fraction before object classification, (b) 
the recovered fraction after.}

\figcaption[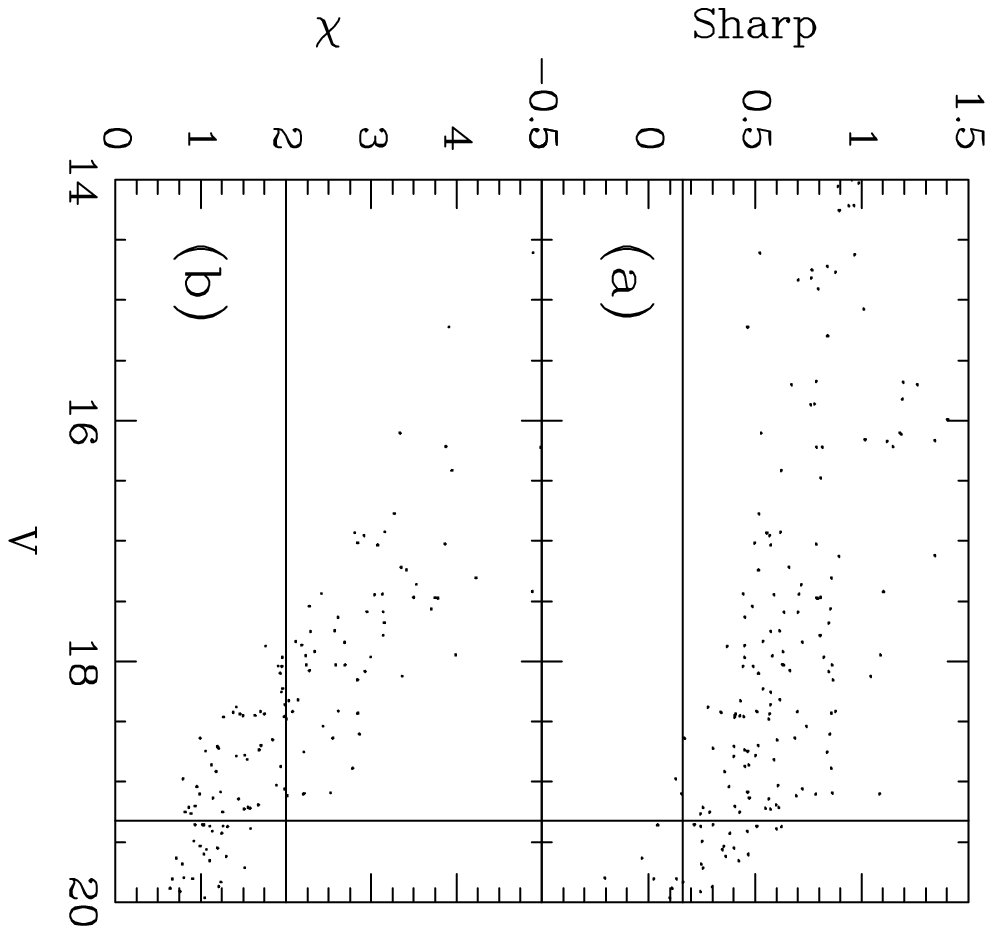]{The magnitude-SHARP and magnitude-$\chi$ distribution of ARTDATA artificial 
galaxies photometered by DAOPHOT.  This can be compared with the distribution of real data seen in Figure 4.}

\figcaption[f10.eps]{The magnitude-$\chi$ and magnitude-SHARP distribution of objects 
photometered in two Swope fields and deeper du Pont imaging.  Open boxes are galaxies measured on both
telescopes, filled circles are stars measured
on both telescopes, dots are stars measured in only one dataset.  These classifications are based on the du Pont
imaging.  Lines represent the cuts made in magnitude, 
SHARP and $\chi$ for classification on the 1 meter data.  The field ASA184-3 had a brighter classification limit and has been shifted 0.6 
magnitudes to align the classification limits.  Note the upward 
slope of the stellar locus where the 2.5 meter data become saturated.}

\figcaption[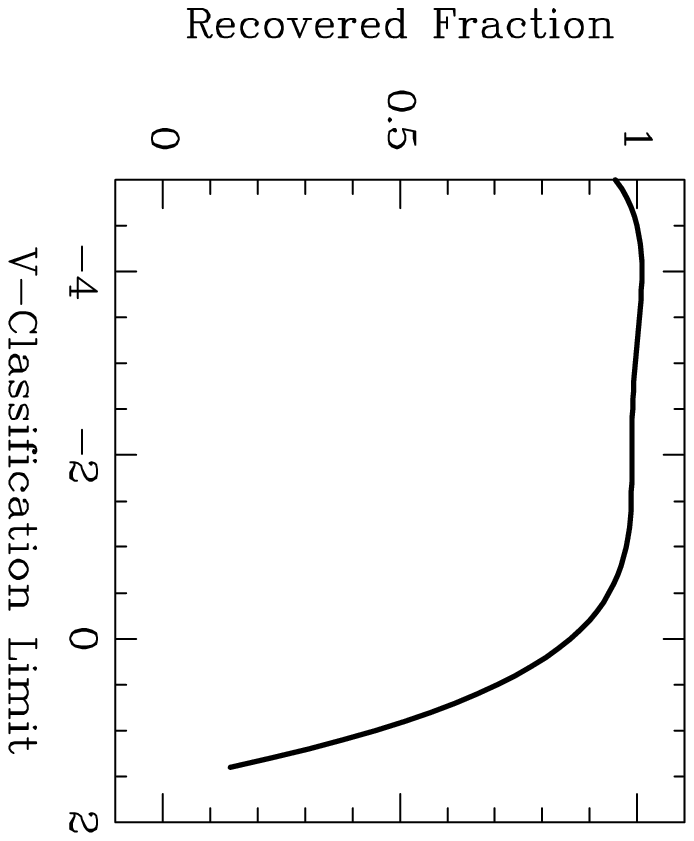]{The recovered fraction of stars in relation to the classification limit of
the Swope images from a comparison of Swope data to deeper du Pont imaging.}

\figcaption[f12.eps]{An illustration of the blue edge of the field stars and the reddening correction.  Panel 
(a) shows the blue edge of the field SA107, a field where it is clearly defined.  The arrows delineate the MSTO of the thick
disk and halo, respectively (see Paper V).  Panel (b) shows the undereddened field SA95, which has highly differential 
reddening.  Panel (c) shows the blue edge of SA95 after reddening correction and demonstrates the much 
tighter blue edge that results.  Note the superiority of the SA107 data (which was primarily obtained 
with the Tek 5 chip) to that of SA95 (entirely Tek 3) in both depth and scatter.}

\figcaption[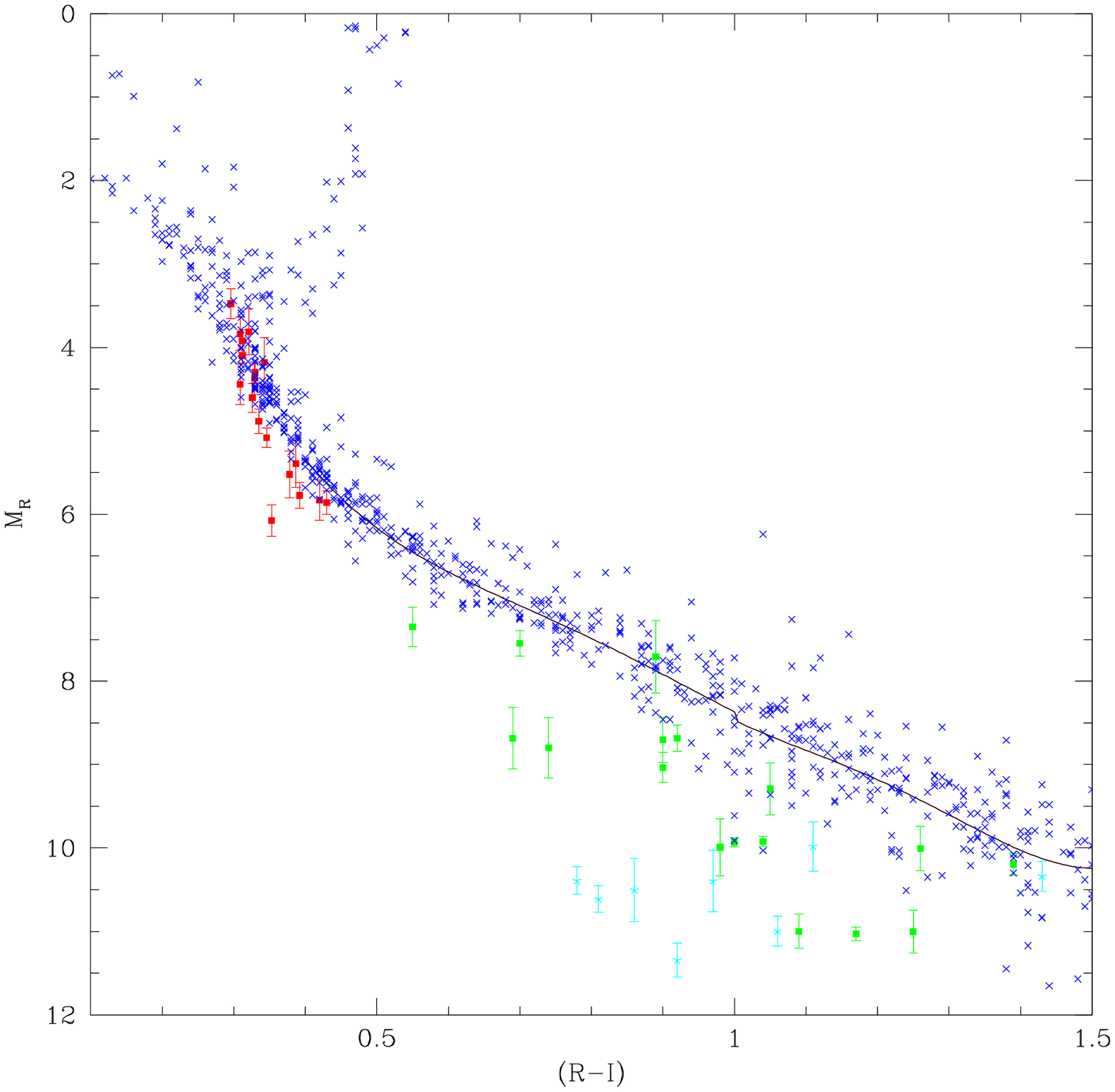]{The absolute magnitude-color distribution of stars in the HIPPARCOS
catalogue used to calibrate photometric parallaxes.  The solid line shows our fit
to the data, excluding potential binaries (the second brighter sequence) and subdwarf stars 
(marked with error bars).  Note the MSTO of the old Galactic populations, which is well blueward
of the $R-I=0.4$ cutoff we use in our evaluation of the density laws.}

\figcaption[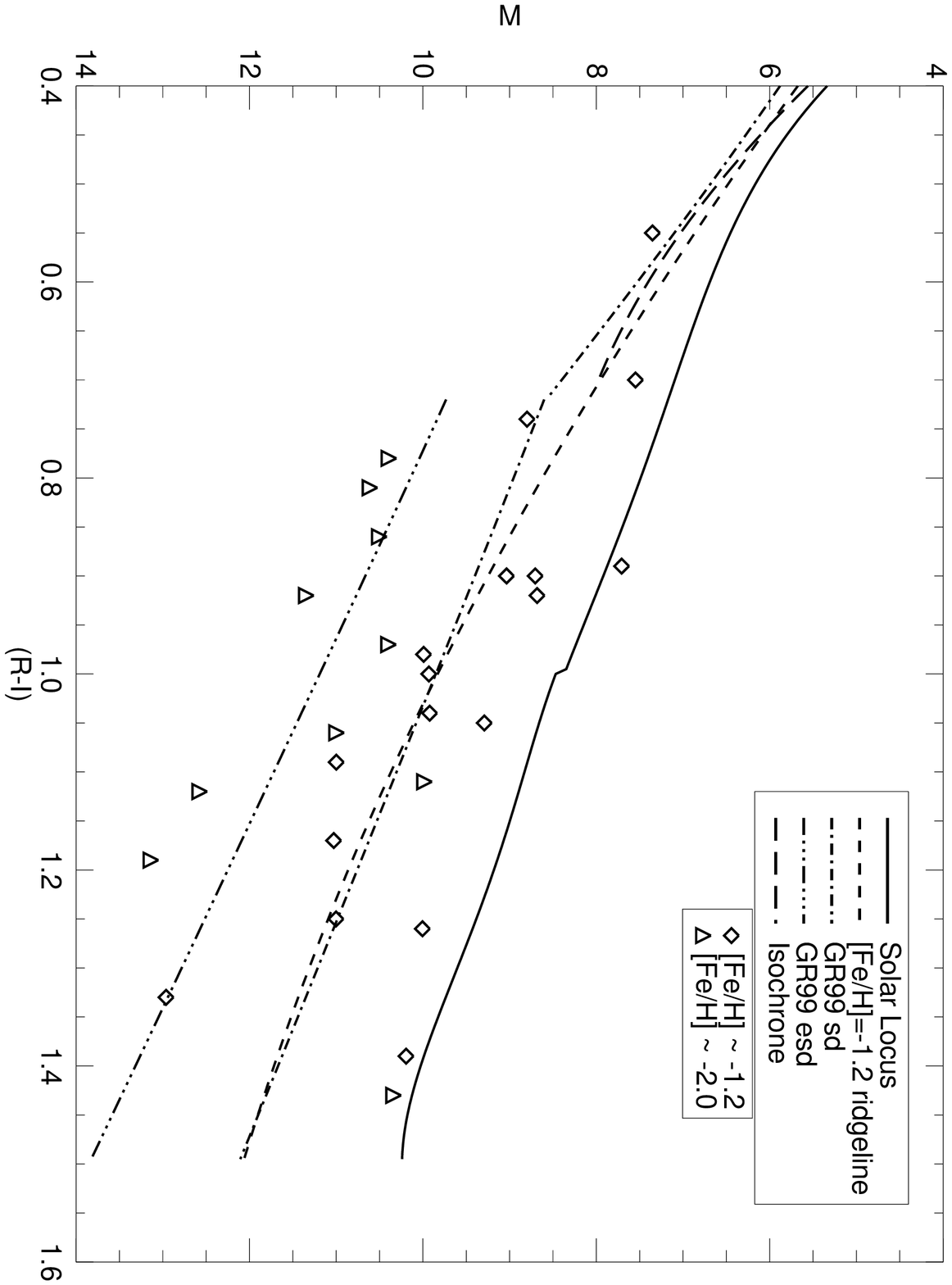]{Our adopted subdwarf correction.  The solid line reflects the H-R diagram shape
we derive at near solar metallicity and the dashed line shows the H-R diagram we derived for [Fe/H]=-1.2.
The H-R curves derived in Gizis \& Reid (1999) for both subdwarfs (``sd", [Fe/H]$\sim-1.2$) and extreme 
subdwarfs (``esd", [Fe/H]$\sim-2.0$) are shown as dot-dash lines.  Also included is 
a fit to the upper main sequence of an [Fe/H]=-1.14 isochrone from BV01 (long dash line), which 
truncates near the top of the main sequence.  The points are metal-poor stars with 
trigonometric parallax taken from Gizis (1997) after application of the Lutz-Kelker correction.}

\figcaption[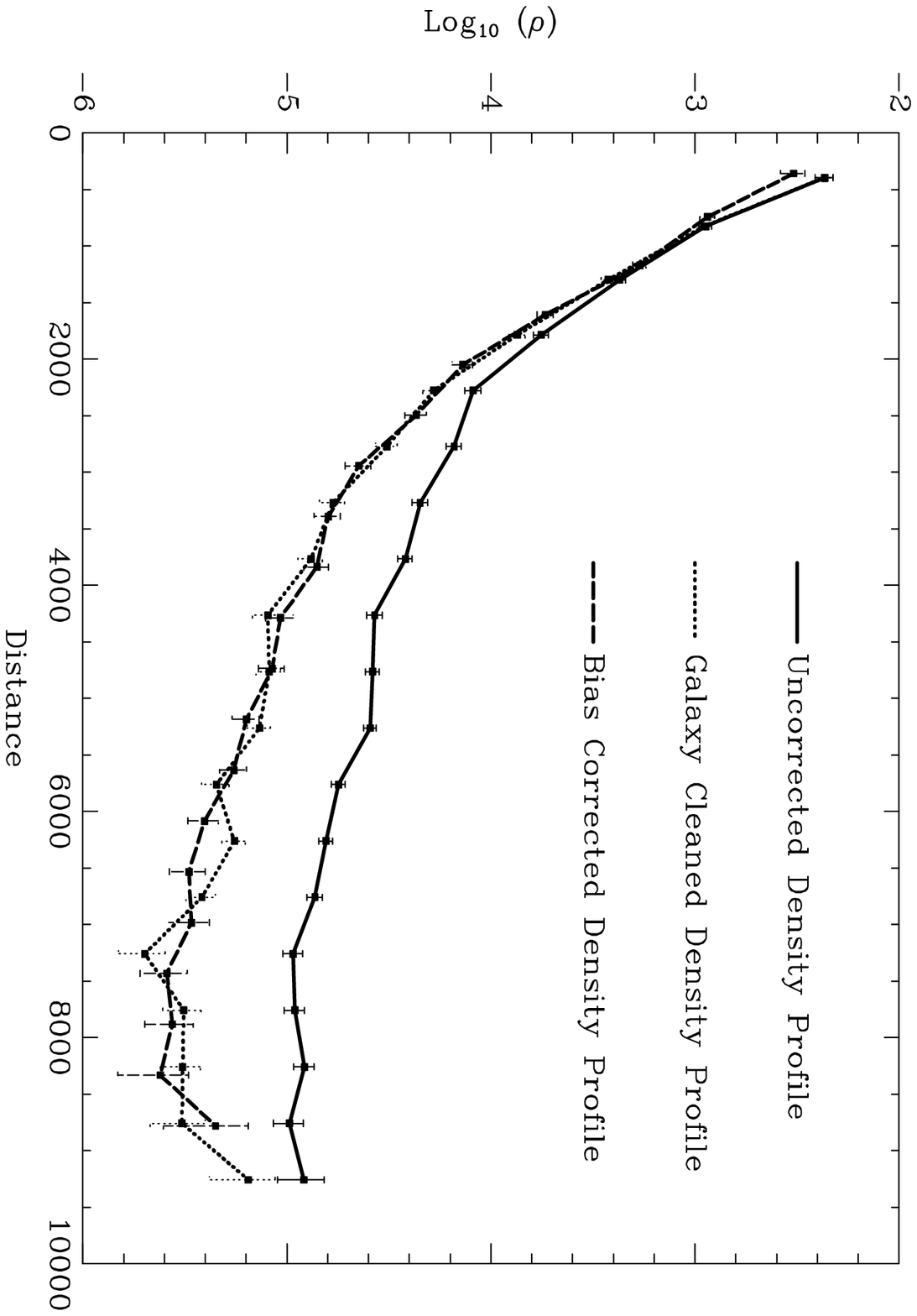]{The effect of Malmquist and subdwarf bias correction and galaxy decontamination on 
the density distribution of stars in SA101.  Note that while removing extragalactic objects substantially
changes the density profile, the Malmquist and subdwarf bias corrections are subtle.  The Malmquist
bias correction shifts the density profile at small distances while the subdwarf correction shortens
the distance range.}

\figcaption[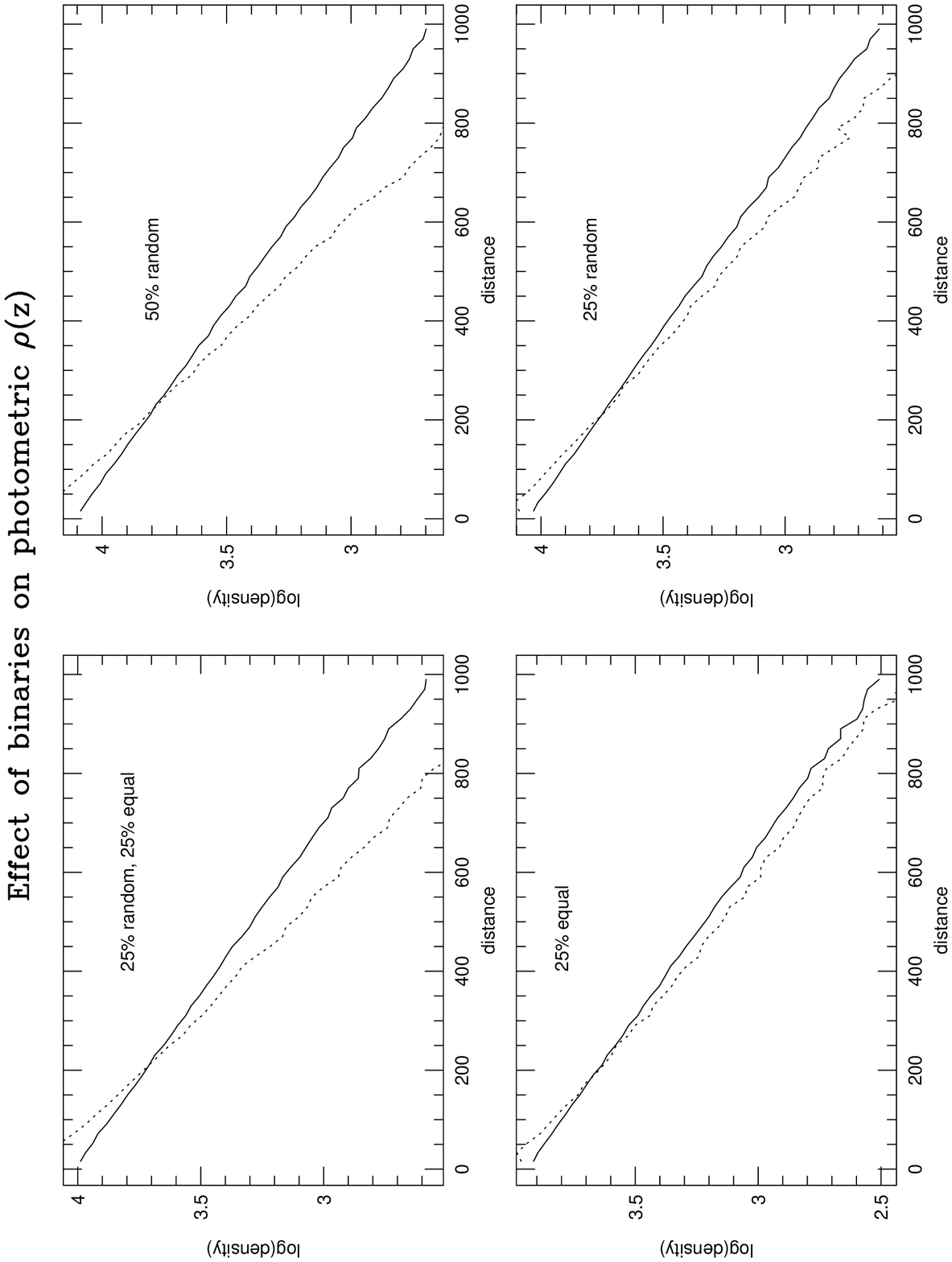]{An evaluation of the effect of binaries upon the derived density law for 
different ratio of near-equal mass and random mass binaries.  The panels
show the actual density law (solid line) against the derived density law (dotted line) for various binary 
fraction.  Note that for each sample, the result is a steepening of the measured density law and 
consequent underestimation of the scale height.}

\figcaption[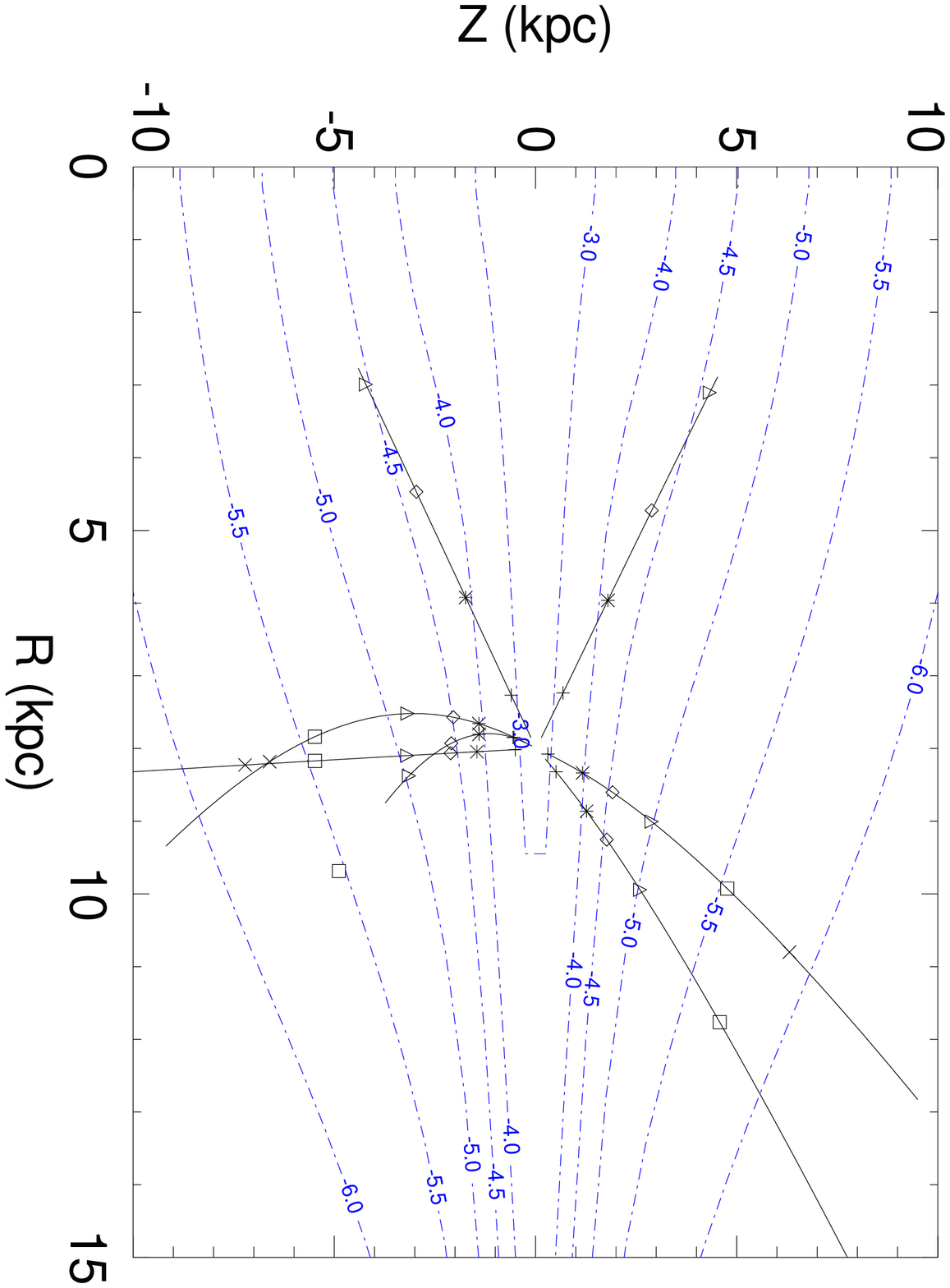]{Comparison of the density isopleths in our bluest stars (symbols) along
the seven lines of sight used in this study (solid lines) 
against the predictions from the Paper I model (dashed lines).  Note the reasonable fit to the SGP but the
strong overpredictions toward the Galactic Center.  Units of density are stars per pc$^{-3}$.  Crosses, 
asterisks, diamonds, triangles, squares and x's are, respectively, densities of -3.0, -4.0, -4.5, -5.0, -5.5,
and -6.0 stars per pc$^{-3}$.}

\figcaption[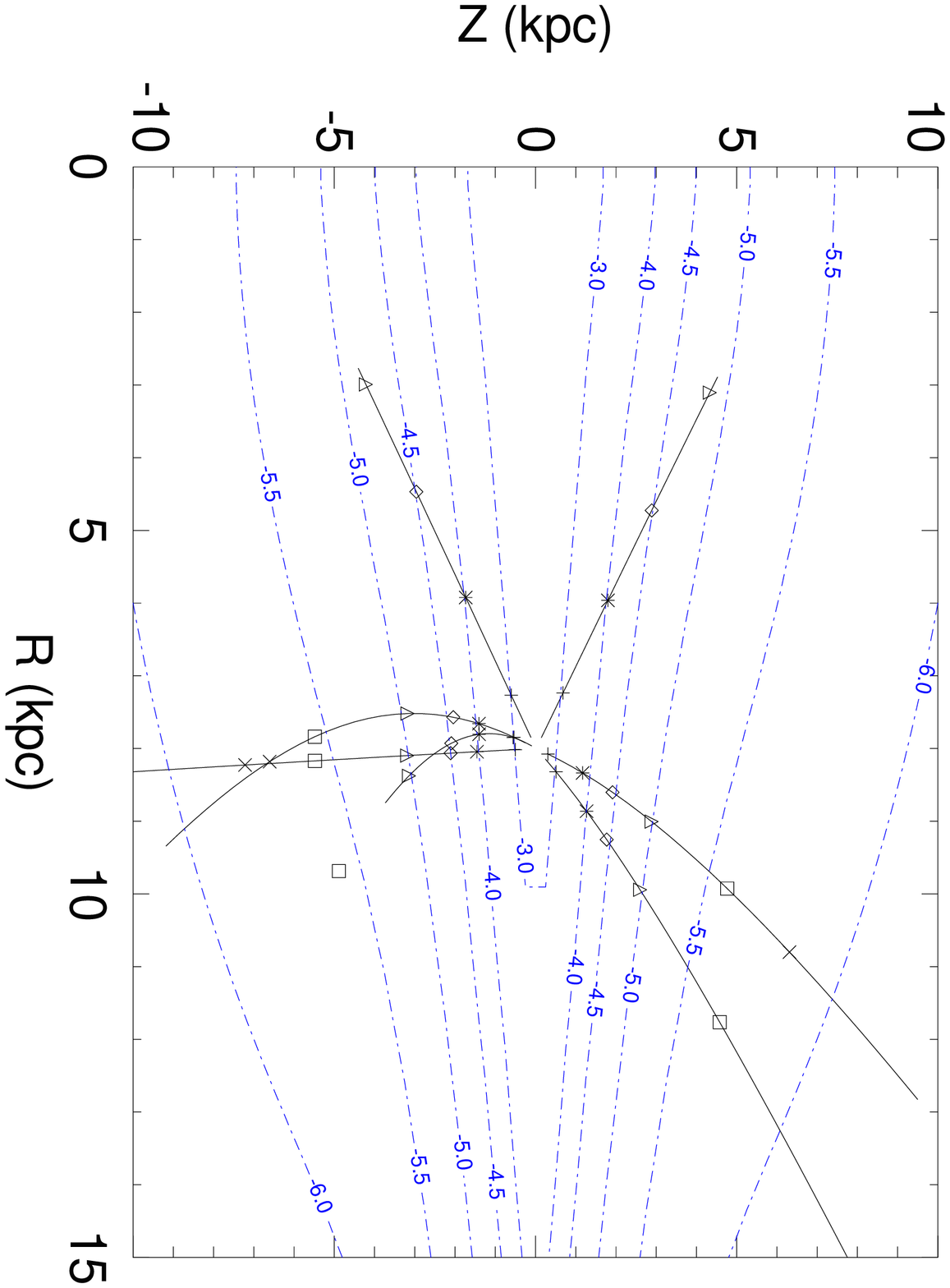]{Comparison of the density isopleths in our bluest stars along the seven lines 
of sight used in this study (solid lines) against the predictions from the model parameterized in Table III 
using the R00 II halo (dashed lines).  Units are identical to figure 17.  Note the excellent fit to 
nearby points but the overpredictions toward the Galactic center and underpredictions in the outer Galaxy.}

\figcaption[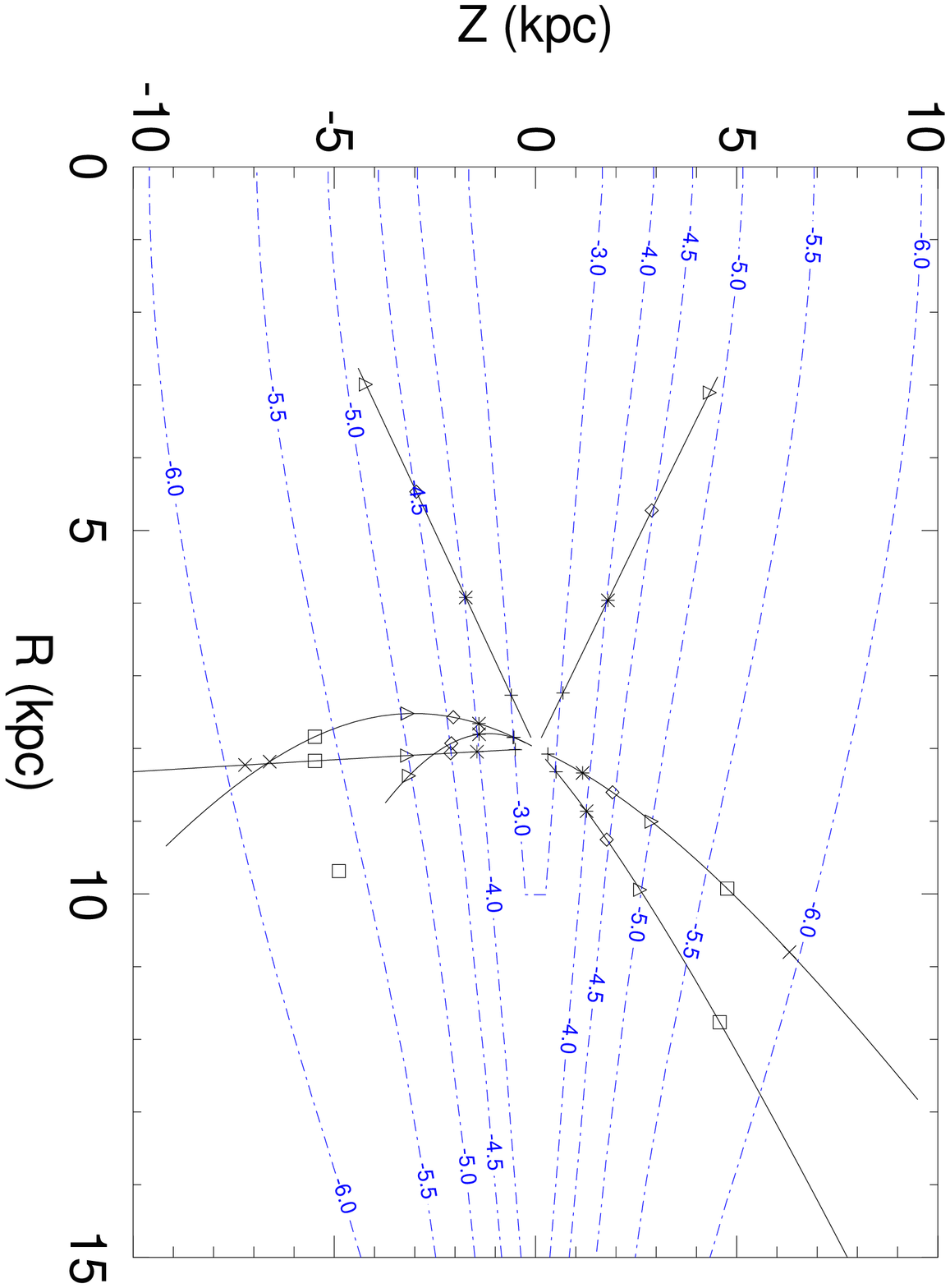]{Comparison of the density isopleths in our bluest stars along the seven lines 
of sight used in this study (solid lines) against the predictions from the model parameterized in Table III
using the L96 halo (dashed lines).  Units are identical to figure 17.  Note the excellent fit to 
nearby points but the overpredictions toward the Galactic center and underpredictions in the outer Galaxy.
Note also that the minor difference between this figure and figure 19.}

\clearpage
\begin{deluxetable}{cccccccccl}
\small
\tablewidth{0 pt}
\tablecaption{Previous Starcount Results}
\tablehead{
\colhead{$Z_{0,thin}$} &
\colhead{$R_{0,thin}$} &
\colhead{$\rho_{thick}$} &
\colhead{$Z_{0,thick}$} &
\colhead{$R_{0,thick}$} &
\colhead{$\rho_{halo}$} &
\colhead{$R_{e,halo}$} &
\colhead{$\frac{c}{a}$} &
\colhead{Reference}\\
\colhead{(pc)} &
\colhead{(kpc)} &
&
\colhead{(kpc)} &
\colhead{(kpc)} &
&
\colhead{(kpc)} &
&}
\startdata
310-325            & \nodata  & .0125-.025        & 1.92-2.39 & \nodata  &
\nodata            & \nodata  & \nodata           & Yoshii 1982\tablenotemark{a} \nl
300                & \nodata  & .02               & 1.45      & \nodata  &
.002               & 3000     & 0.85              & GR83 \nl
325                & \nodata  & .02               & 1.3       & \nodata &
.002               & 3000     & 0.85              & Gilmore 1984 \nl
280                & \nodata  & .0028             & 1.9       & \nodata  &
.0012              & \nodata  &  \nodata          & Tritton \& Morton 1984 \nl
200-475            & \nodata  & .016              & 1.18-2.21 & \nodata &
.0016              & \nodata  & 0.8               & Robin \& Creze 1986 \nl
300                & \nodata  & .02               & 1.0       & \nodata  &
.001               & \nodata  & 0.85              & del Rio \& Fenkart 1987 \nl
285                & \nodata  & .015              & 1.3-15    & \nodata  &
.002               & 2360     & flat              & Fenkart et al. 1987 \nl
325                & \nodata  & .0224             & .95       & \nodata  &
.001               & 2.9      & 0.9               & Yoshii et al. 1987 \nl
249                & \nodata  & .041              & 1.0       & \nodata &
.002               & 3000     & 0.85              & Kuijken \& Gilmore 1989 \nl
350                & 3.8      & .019              & .9        & 3.8      &
.0011              & 2.7      & 0.84              & Yamagata \& Yoshii 1992\nl
290                & \nodata  & \nodata           & 0.86      & \nodata  &
\nodata            & 4.0      & \nodata           & von Hippel \& Bothun 1992\nl
325                & \nodata  & .0225             & 1.5       & \nodata  &
.0015              & 3.5      & 0.80              & Paper I\nl
325                & 3.2      & .019              & 0.98      & 4.3      &
.0024              & 3.3      & 0.48              & L96\nl
\nodata            & 2.5      & .056              & 0.76      & 2.8      &
.0015              & 2.44-2.75\tablenotemark{b}  & 0.6-0.85          & R96, R00\nl
290                & 4.0      & .059              & 0.91      & 3.0      & 
.0005              & 2.69     & 0.84              & Buser et al. 1998, B99\nl
240                & 2.5      & .061              & .79       & 2.8      &
\nodata            & \nodata  & 0.6-0.85          & Ojha et al. 1999 \nl
330                & 2.250    & .065-.13          & 0.58-0.75 & 3.5      &
.00125             & \nodata  & 0.55              & Chen et al. 2001 \nl
\enddata
\tablenotetext{a}{The stars Yoshii 1982 assigned to the halo have been here assigned to the thick
disk.}
\tablenotetext{b}{Power Law Index replacing $R_e$.}
\end{deluxetable}

\clearpage
\begin{deluxetable}{ccc}
\tablewidth{0 pt}
\tablecaption{Program Fields}
\tablehead{
\colhead{Selected Area} &
\colhead{$(\alpha,\delta)_{1950.0}$} &
\colhead{$(l,b)$}}
\startdata
141   & 01:05.4,-29:34 & 245,-86 \nl
118   &  02:17.4,-14:36 & 186,-66 \nl
95    &  03:52.6,+00:09 & 189,-38 \nl
96    &  04:50.6,+00:05 & 198,-26 \nl
194-18&  05:09.5,+07:37 & 194,-18 \nl
101   & 09:54.6,-00:14  & 239,+40 \nl
102   & 10:51.0,-0:49   & 253,+50 \nl
104   &  12:40.6,-00:16 & 299,+62 \nl
107   & 15:36.6,-00:10  & 6,+41   \nl
184   & 20:52.4,-44:39 & 356,-40 \nl
90    & 22:14.4,+15:25  & 77,-33  \nl
114   &  22:39.6,+00:26 & 70,-48 \nl
\enddata
\end{deluxetable}

\begin{deluxetable}{cccccccc}
\tablewidth{0 pt}
\tablecaption{Halo Model Comparisons to Intrinsically Bright ($5.8 \leq R-I < 6.8$) Stars}
\tablehead{
\colhead{$\rho_{thin}$} &
\colhead{$Z_{thin}$} &
\colhead{$R_{thin}$} &
\colhead{$\rho_{thick}$} &
\colhead{$Z_{0,thick}$} &
\colhead{$R_{0,thick}$} &
\colhead{$\rho_{halo}$} &
\colhead{$\chi^2$}\\
\colhead{$(pc^{-3}$)} &
\colhead{$(pc)$} &
\colhead{$(pc)$} &
\colhead{$(\rho_{thin}^{-1})$} &
\colhead{$(pc)$} &
\colhead{$(pc)$} &
\colhead{$(\rho_{thin}^{-1})$} &
\colhead{}}
\startdata
\multicolumn{8}{c}{Exponential Disks, L96 halo ($\frac{c}{a}=0.48, R_e=3.28$ kpc)}\\
$0.0048 \pm .0012$ & 280          & $2200^{+14000}_{-800}$ & 
$0.080 \pm .020$   & $710 \pm 60$ & $3800^{+4800}_{-1000}$ & 
$.0024 \pm .0010 $ & 2.61\nl
\multicolumn{8}{c}{Exponential Disks, R00 halo I ($\frac{c}{a}=.76, n=2.44$)}\\
$0.0047 \pm .0012$ & 280          & $2000 \pm^{+7400}_{-700}$  & 
$0.081 \pm .017$   & $770 \pm 60$ & $4000 \pm^{+3100}_{-1000}$ & 
$.0008 \pm .0005$  & 2.84\nl
\multicolumn{8}{c}{Exponential Disks, R00 halo II ($\frac{c}{a}=.60, n=2.75$)}\\
$0.0047 \pm .0012$ & 280          & $2100^{+10000}_{-700}$  & 
$0.084 \pm .0018$  & $740 \pm 50$ & $3900^{+3400}_{-1000}$ & 
$.0013 \pm .0007 $ & 2.70\nl
\multicolumn{8}{c}{Exponential Disks, N97 halo ($\frac{c}{a}=1.0, n=3.0$)}\\
$0.0046 \pm .0011$ & 280          & $1700 \pm^{+2500}_{-500}$  & 
$0.070 \pm .014$   & $850 \pm 60$ & $4800 \pm^{+4900}_{-1300}$ & 
$.0004 \pm .0003 $ & 3.28\nl
\multicolumn{8}{c}{Exponential Disks, B99 halo ($\frac{c}{a}=.85, R_e=2.69$ kpc)}\\
$0.0047 \pm .0011$ & 280          & $1800 \pm^{+3200}_{-500}$  & 
$0.068 \pm .013$   & $860 \pm 60$ & $4800 \pm^{+4500}_{-1300}$ & 
$.0004 \pm .0004$  & 3.25\nl
\hline
\enddata
\tablenotetext{}{Scaleheights not corrected for binarism.}
\end{deluxetable}

\begin{deluxetable}{lccccccc}
\tablewidth{0 pt}
\tablecaption{Best Fit for Exponential Disks, Flattened $R_G^{-2.75}$ Halo}
\tablehead{
\colhead{Cut} &
\colhead{$\rho_{thin}$} &
\colhead{$Z_{0,thin}$} &
\colhead{$R_{0,thin}$} &
\colhead{$\rho_{thick}$} &
\colhead{$Z_{0,thick}$} &
\colhead{$R_{0,thick}$} &
\colhead{$\chi^2$} \\
\colhead{} &
\colhead{$(pc^{-3}$)} &
\colhead{$(pc)$} &
\colhead{$(pc)$} &
\colhead{$(\rho_{thin}^{-1})$} &
\colhead{$(pc)$} &
\colhead{$(pc)$} &
\colhead{}}
\startdata
$5.8 \leq R-I < 6.8$     & $0.0047 \pm .0012$ & $280 \pm 20$             & $2100^{+10000}_{-700}$ 
                         & $0.084 \pm .018$   & $740 \pm 50$             & $3900^{+3400}_{-1000}$ & 2.70\nl
%.0013 halo
$6.8 \leq M_R < 7.8$     & $0.0046 \pm .0010$ & $275 \pm 20$             & $2300^{+8000}_{-900}$ 
                         & $0.083 \pm .013$   & $900 \pm 70$             & $3600^{+1800}_{-800}$  & 1.91\nl
%.0015 halo
$7.8 \leq M_R < 8.8$     & $0.0058 \pm .0008$ & $295 \pm 15$             & $2400^{+2100}_{-700}$ 
                         & $0.027 \pm .007$   & $1560^{+420}_{-330}$     & $2400^{+2100}_{-700}$  & 1.83\nl
$8.8 \leq M_R \leq 10.2$ & $0.0143 \pm .0009$ & $355 \pm 10$             & $2350^{+1800}_{-500}$  
                         & $0.013 \pm .009$   & $6200^{+\infty}_{-5500}$ & $7000^{+\infty}_{-6000}$ & 2.18\nl
\enddata
\tablenotetext{}{Scaleheights not corrected for binarism.}
\end{deluxetable}

\begin{deluxetable}{lccccccc}
\tablewidth{0 pt}
\tablecaption{Best Fit for $sech^2$ Disks, Flattened $R_G^{-2.75}$ Halo}
\tablehead{
\colhead{Cut} &
\colhead{$\rho_{thin}$} &
\colhead{$Z_{0,thin}$} &
\colhead{$R_{0,thin}$} &
\colhead{$\rho_{thick}$} &
\colhead{$Z_{0,thick}$} &
\colhead{$R_{0,thick}$} &
\colhead{$\chi^2$} \\
\colhead{} &
\colhead{$(pc^{-3}$)} &
\colhead{$(pc)$} &
\colhead{$(pc)$} &
\colhead{$(\rho_{thin}^{-1})$} &
\colhead{$(pc)$} &
\colhead{$(pc)$} &
\colhead{}}
\startdata
$5.8 \leq M_R < 6.8$     & $0.0025 \pm .0007$ & $230 \pm 20$          & $3700^{+infty}_{-2200}$
                         & $0.097 \pm .020$   & $570 \pm 40$          & $3000^{+1700}_{-600}$     & 2.73\nl
%rhohalo=.0035
$6.8 \leq M_R < 7.8$     & $0.0023 \pm .0005$ & $240 \pm 20$             & $2700^{+\infty}_{-1100}$ 
                         & $0.079 \pm .011$   & $780 \pm 50$             & $3600^{+1500}_{-700}$    & 2.28\nl
$7.8 \leq M_R < 8.8$     & $0.0032 \pm .0005$ & $230 \pm 10$             & $2600^{+4800}_{-1000}$ 
                         & $0.053 \pm .009$   & $860 \pm 110$            & $2400^{+1200}_{-500}$    & 1.59\nl
$8.8 \leq M_R \leq 10.2$ & $0.0085 \pm .0006$ & $255 \pm 10$             & $2600^{+1400}_{-800}$  
                         & $0.045 \pm .012$   & $3800^{+\infty}_{-3000}$ & $2500^{+\infty}_{-1200}$ & 1.89\nl
\enddata
\tablenotetext{}{Scaleheights not corrected for binarism.}
\end{deluxetable}

\begin{deluxetable}{ccc}
\tablewidth{0 pt}
\tablecaption{The New Model}
\tablehead{
\colhead{Parameter} &
\colhead{Value} & 
\colhead{Corrected Value\tablenotemark{a}}}
\startdata
$Z_{0,thin}$\tablenotemark{b}   & 280 pc & 350 pc \nl
$Z_{0,thin}$\tablenotemark{c}   & 230 pc & 290 pc\nl
$R_{0,thin}$   & 2-2.5 kpc & \nl
$\rho_{thick}$ & 6-10\% & \nl
$Z_{0,thick}$\tablenotemark{b}  & 700-1000 pc & 900-1200 pc\nl
$Z_{0,thick}$\tablenotemark{c}  & 500-800 pc  & 600-1000 pc\nl
$R_{0,thick}$  & 3-4 kpc & \nl
$\rho_{halo}$  & .15\% & \nl
$\frac{c}{a}$  & 0.5-0.7 & \nl
\enddata
\tablenotetext{a}{Scaleheights corrected for binarism with 50\% binary fraction.}
\tablenotetext{b}{Exponential Scaleheight}
\tablenotetext{c}{$sech^2$ Scaleheight}
\end{deluxetable}

\clearpage
\begin{figure}[h]
\plotone{f1.eps}
\end{figure}

\begin{figure}[h]
\plotone{f2.eps}
\end{figure}

\begin{figure}[h]
\plotone{f3.eps}
\end{figure}

\clearpage

\begin{figure}[h]
\plotone{f7.eps}
\end{figure}

\begin{figure}[h]
\plotone{f8.eps}
\end{figure}

\begin{figure}[h]
\plotone{f9.eps}
\end{figure}

\begin{figure}[h]
\plotone{f11.eps}
\end{figure}

\begin{figure}[h]
\plotone{f13.eps}
\end{figure}

\begin{figure}[h]
\plotone{f14.eps}
\end{figure}

\begin{figure}[h]
\plotone{f15.eps}
\end{figure}

\begin{figure}[h]
\plotone{f16.eps}
\end{figure}

\begin{figure}[h]
\plotone{f17.eps}
\end{figure}

\begin{figure}[h]
\plotone{f18.eps}
\end{figure}

\begin{figure}[h]
\plotone{f19.eps}
\end{figure}

\end{document}